\documentclass[aps,twocolumn,showpacs,preprintnumbers,floats]{revtex4}
\usepackage{mathrsfs}
\usepackage{longtable,lscape}
\usepackage{txfonts}
\usepackage{amssymb}
\usepackage{indentfirst}
\usepackage{graphicx,,booktabs}
\usepackage{color}
\usepackage{amssymb}
\usepackage{epsfig}
\newcommand{\vsig}{\mbox{\boldmath$\sigma$\unboldmath}}

\begin{document}

\title{Low energy reactions $K^-p\rightarrow\Sigma^0\pi^0$, $\Lambda\pi^0$, $\bar{K}^0n$ and the strangeness $S=-1$ hyperons}
\author{
Xian-Hui Zhong$^{1,3}$ \footnote {zhongxh@hunnu.edu.cn} and Qiang
Zhao$^{2,3}$ \footnote {zhaoq@ihep.ac.cn}}

\affiliation{ 1) Department of Physics, Hunan Normal University, and
Key Laboratory of Low-Dimensional Quantum Structures and Quantum
Control of Ministry of Education, Changsha 410081, China }

\affiliation{ 2) Institute of High Energy Physics,
       Chinese Academy of Sciences, Beijing 100049, China
}

\affiliation{ 3) Theoretical Physics Center for Science Facilities,
Chinese Academy of Sciences, Beijing 100049, China }


\begin{abstract}
A combined study of the reactions $K^-p\rightarrow \Sigma^0\pi^0$,
$\Lambda\pi^0$ and $\bar{K}^0n$ at low energies is carried out with
a chiral quark-model approach. Good descriptions of the experimental
observations are obtained. The roles of the low-lying strangeness
$S=-1$ hyperon resonances in these processes are carefully analyzed.
We find that: (i) In the $K^-p\rightarrow \Sigma^0\pi^0$ process,
both $\Lambda(1405)S_{01}$ and $\Lambda(1520)D_{03}$ play dominant
roles. Significant contributions of $\Lambda(1670)S_{01}$ and
$\Lambda(1690)D_{03}$ could be seen around their threshold; (ii) In
the $K^-p\rightarrow \Lambda\pi^0$ process, some obvious evidence of
$\Sigma(1775)D_{15}$ and $\Sigma(1750)S_{11}$ could be found. Some
hints of $\Sigma(1620)S_{11}$ might exist in the reaction as well.
$\Sigma(1750)S_{11}$ and $\Sigma(1620)S_{11}$ should correspond to
the representations $[70,^48]S_{11}$ and $[70,^28]S_{11}$,
respectively; (iii) In the $K^-p\rightarrow \bar{K}^0n$ process, the
dominant resonances are $\Lambda(1405)$ and $\Lambda(1520)$. Some
evidence of $\Lambda(1690)D_{03}$, $\Sigma(1670)D_{13}$ and
$\Sigma(1775)D_{15}$ could be seen as well. A weak coupling of
$\Lambda(1670)S_{01}$ to $\bar{K}N$ should be needed in the
reactions $K^-p\rightarrow \Sigma^0\pi^0$ and $\bar{K}^0n$.
Furthermore, by analyzing these reactions, we also find that the
$u$-, $t$-channel backgrounds and $s$-channel Born term play crucial
roles in the reactions: (i) The angle distributions of
$K^-p\rightarrow \Sigma^0\pi^0$ are very sensitive to the $u$-,
$t$-channel backgrounds and s channel $\Lambda$ pole; (ii) The
reaction $K^-p\rightarrow\Lambda \pi^0$ is dominated by the $u$-,
$t$-channel backgrounds and the ground $P$-wave state
$\Sigma(1385)P_{13}$; (iii) While, the reaction $K^-p\rightarrow
\bar{K}^0 n$ is governed by the $t$-channel background, and
$\Sigma(1385)P_{13}$ also plays an important role in this reaction.

\end{abstract}


\maketitle

\section{Introduction}

There exist many puzzles in the spectroscopies of $\Lambda$ and
$\Sigma$ hyperons. To clearly see the status of these hyperon
spectroscopies, we have collected all the strangeness $S=-1$
hyperons classified in the quark model up to $n=2$ shell in Tab.
\ref{Resonance}. From the table, it is seen that only a few
low-lying $S=-1$ hyperons are established, while for most of them
there is still no confirmed evidence found in experiments.
Concretely, for the $\Sigma$ spectroscopy although a lot of states,
such as $\Sigma(1480)\textbf{Bumps}$, $\Sigma(1560)\textbf{Bumps}$,
$\Sigma(1670)\textbf{Bumps}$ and $\Sigma(1690)\textbf{Bumps}$, have
been listed by the Particle Data Group (PDG) \cite{PDG}, they are
not established at all. Their quantum numbers and structures are
still unknown. Even for the well-established states with known
quantum numbers, such as $\Sigma(1750)1/2^-$, it is questionable in
the classification of them according to various quark
models~\cite{Chen:2009de,Klempt:2009pi,Melde:2007,Melde:2008,Isgur78}.
For the $\Lambda$ spectroscopy, a little more knowledge is known
compared with that of $\Sigma$, however, the properties of some
$\Lambda$ resonances with confirmed quantum numbers are still
controversial. For example, it is still undetermined whether these
states, such as $\Lambda(1405)$, $\Lambda(1670)$ and
$\Lambda(1520)$, are excited three quark states or dynamically
generated resonances, though their $J^P$ are
well-determined~\cite{Klempt:2009pi}. How to clarify these issues
and extract information of the unestablished hyperon resonances from
experimental data are still open questions.

To uncover the puzzles in the $S=-1$ hyperon spectroscopies, many
theoretical and experimental efforts have been performed.
Theoretically, (i) the mass spectroscopes were predicted in various
quark models~\cite{Isgur78,Isgur:1977ky,Capstick:1986bm,Melde:2008,
Gerasyuta:2007, Bijker:2000,
Glozman:1997ag,Loring:2001ky,Schat:2001xr,Goity:2002pu}, large $N_c$
QCD approach~\cite{Schat:2001xr,Goity:2002pu} and lattice QCD
etc~\cite{Menadue:2011pd,Engel:2012qp}; (ii) the strong decays were
studied within different
models~\cite{Koniuk:1979vy,Melde:2006yw,Melde:2008,Melde:2007,An:2010wb};
(iii) the properties of the individual resonances, such as
$\Lambda(1405)$, $\Lambda(1670)$ and $\Lambda(1520)$, were attempted
to extract from the $K^-p$ scattering data with U$\chi$PT approaches
~\cite{Hyodo:2011ur,Oller:2005ig,Roca:2006sz,Oller:2006hx,Borasoy:2005ie,Hyodo:2003qa,Oset:2001cn,
GarciaRecio:2002td,Jido:2003cb,Oller:2000fj,Oset:1997it,Oller:2006jw,Borasoy:2006sr,Roca:2008kr},
B$\chi$PT approaches \cite{Bouzasa;2008}, $K$-matrix methods
\cite{Martin:1969ud,D. M. Manley}, large-$N_c$ QCD method
\cite{Lutz:2001yb}, meson-exchange models
\cite{Buttgen:1985yz,Buettgen:1990yw,MuellerGroeling:1990cw}, quark
model approaches \cite{Hamaie:1995wy,Zhong:2009}, dispersion
relations \cite{Gensini:1997fp,Martin:1980qe}, and the other
hadronic models \cite{BS:2009,Gao:2012zh,Gao:2010ve,Liu:2011sw};
(iv) furthermore, the possible exotic properties of some strange
baryons, such as two mesons-one baryon bound states, quark mass
dependence, five-quark components, were also discussed in the
literature \cite{Mart:2007x, GarciaRecio:2003ks,Zou:2008be}.
Experimentally, the information of the hyperon resonances was mainly
obtained from the measurements of the reactions $\bar{K}N\rightarrow
\bar{K}N$, $\Sigma \pi$, $\Lambda\pi$, $\eta n$, $\Sigma \pi\pi$ and
$\Lambda \pi\pi$
\cite{Mast:1976,Baxter:1974zs,Ponte:1975bt,Conforto:1975nw,Evans:1983hz,
Kim:1965,Jones:1974at,AlstonGarnjost:1977cs,AlstonGarnjost:1977ct,
Cameron:1980nv,Ciborowski:1982et,armen:1970zh,London:1975av,Olmsted:2004,
Mast:1974sx,Berley:1996zh,Bangerter:1980px,Starostin:2001zz,Prakhov:2004an,
Zychor:2007gf}. In recent years, some other new experiments, such as
excited hyperon productions from $\gamma N$ and $NN$ collisions, had
been carried out at LEPS, JLAB and COSY to investigate the hyperon
properties further~\cite{Niiyama:2008rt,Kohri:2009xe,
Zychor:2008ct,Zychor:2006}.

\begin{table}[ht]
\begin{center}
\caption{ The classification of the strangeness $S=-1$ hyperons in
the quark model up to $n=2$ shell. The ``$?$" denotes a resonance
being unestablished. $l_{I,2J}$ is the PDG notation of baryons.
$N_6$ and $N_3$ denote the SU(6) and SU(3) representation,
respectively. $\textbf{L}$ and $\textbf{S}$ stand for the total
orbital momentum and spin of the baryon wave function,
respectively.} \label{Resonance}
\begin{tabular}{|c|c|c|c|}\hline\hline
  $[N_6,^{2\mathbf{S}+1}N_3,n,\mathbf{L}]$ & $l_{I,2J}$ & $l_{I,2J}$
                    \\ \hline
  $[\textbf{56},^2\textbf{8},0,\mathbf{0}]$ & $P_{01}(1116)$   &
$P_{11}(1193)$
                     \\
 $[\textbf{56},^4\textbf{10},0,\mathbf{0}]$ &   ...        &
$P_{13}(1385)$\\ \hline
  $[\textbf{70},^2\textbf{1 },1,1]$  & $S_{01}(1405)$   &...\\
  & $D_{03}(1520)$   &...\\
    \hline
   $[\textbf{70},^2\textbf{10 },1,\mathbf{1}]$ &   ...         & $S_{11}(?)$   \\
  &     ...       & $D_{13}(?)$  \\
    \hline

   $[\textbf{70},^2\textbf{8 },1,\mathbf{1}]$ & $S_{01}(1670)$  & $S_{11}(?)$   \\
   & $D_{03}(1690)$  & $D_{13}(1670)$  \\
    \hline

   $[\textbf{70},^4\textbf{8 },1,\mathbf{1}]$ & $S_{01}(1800)$  & $S_{11}(?)$   \\
     & $D_{03}(?)$     & $D_{13}(?)$  \\
      & $D_{05}(1830)$     & $D_{15}(1775)$  \\
\hline\hline

  $[\textbf{56},^2\textbf{8},2,\mathbf{0}]$ & $P_{01}(1600)$   &
$P_{11}(1660)$\\ \hline $[\textbf{56},^2\textbf{8},2,\mathbf{2}]$ &
$P_{03}(?)$
& $P_{13}(?)$\\
 & $F_{05}(?)$
& $F_{15}(?)$\\
\hline

  $[\textbf{56},^4\textbf{10},2,\mathbf{0}]$ &    ...       & $P_{13}(?)$\\
\hline
 $[\textbf{56},^4\textbf{10},2,\mathbf{2}]$ &   ...        & $P_{11}(?)$\\
   &    ...       & $P_{13}(?)$\\
   &    ...       & $F_{15}(?)$\\
   &    ...       & $F_{17}(?)$\\
\hline
  $[\textbf{70},^2\textbf{1 },2,\mathbf{0}]$  & $P_{01}(1810?)$   &...\\
\hline
 $[\textbf{70},^2\textbf{1 },2,\mathbf{2}]$  & $P_{03}(?)$   &...\\
    & $F_{05}(?)$   &...\\
\hline
   $[\textbf{70},^2\textbf{10 },2,\mathbf{0}]$ &  ...          & $P_{11}(?)$   \\
\hline
   $[\textbf{70},^2\textbf{10 },2,\mathbf{2}]$ &  ...          & $P_{13}(?)$   \\
   &    ...        & $F_{15}(?)$   \\
\hline
   $[\textbf{70},^2\textbf{8 },2,\mathbf{0}]$ & $P_{01}(?)$  & $P_{11}(?)$   \\
\hline
   $[\textbf{70},^2\textbf{8 },2,\mathbf{2}]$ & $P_{03}(?)$  & $P_{13}(?)$   \\
     & $F_{05}(?)$  & $F_{15}(?)$   \\
\hline
   $[\textbf{70},^4\textbf{8 },2,\mathbf{0}]$ & $P_{03}(?)$  & $P_{13}(?)$   \\
\hline
   $[\textbf{70},^4\textbf{8 },2,\mathbf{2}]$ & $P_{01}(?)$  & $P_{11}(?)$   \\
    & $P_{03}(?)$  & $P_{13}(?)$   \\
     & $F_{05}(?)$  & $F_{15}(?)$   \\
     & $F_{07}(?)$ & $F_{17}(?)$
\\      \hline
\end{tabular}
\end{center}
\end{table}

Recently, some higher precision data of the reactions
$K^-p\rightarrow \Sigma^0\pi^0$
\cite{Manweiler:2008zz,Prakhov:2008}, $\Lambda\pi^0$
 and $\bar{K}^0n$ \cite{Prakhov:2008} at eight momentum beams between
514 and 750 MeV/c were reported, which provides us a good
opportunity to study these low-lying $\Lambda$ and $\Sigma$
resonances systemically. In this work, we carry out a combined study
of these reactions in a chiral quark model, where an effective
chiral Lagrangian is introduced to account for the quark-meson
coupling. Since the quark-meson coupling is invariant under the
chiral transformation, some of the low-energy properties of QCD are
retained. The chiral quark model has been well developed and widely
applied to meson photoproduction
reactions~\cite{qk1,qk2,qkk,Li:1997gda,zhao-kstar,qk3,qk4,qk5,He:2008ty,Saghai:2001yd,Zhong:2011ht}.
Its recent extension to describe the process of $\pi N$
~\cite{Zhong:2007fx} and $\bar{K} N$ ~\cite{Zhong:2009} scattering,
and the charmed hadron strong decays
~\cite{Zhong:2007gp,Zhong:2010vq,Liu:2012sj} also turns out to be
successful and inspiring.

This work is organized as follows. In  Sec.\ \ref{suc}, the
formulism of the model is reviewed. Then, the partial wave
amplitudes are separated in Sec.\ \ref{sres}. The numerical results
are presented and discussed in Sec.\ \ref{MP}. Finally, a summary is
given in Sec.\ \ref{SUM}.

\begin{figure}[ht]
\centering \epsfxsize=8.6 cm \epsfbox{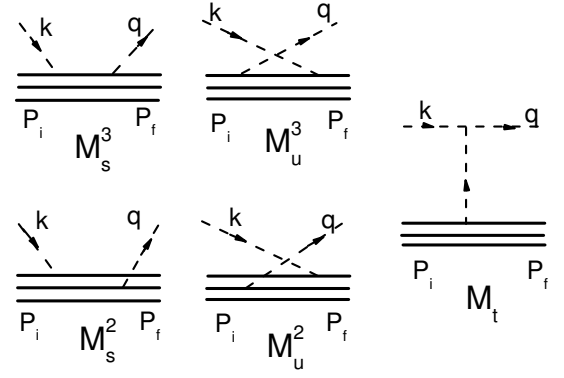}\caption{ $s$, $u$
and $t$ channels are considered in this work. $M^{3}_{s}$ and
$M^{3}_{u}$ ($M^{2}_{s}$, $M^{2}_{u}$) correspond to the amplitudes
of $s$ and $u$ channels for the incoming and outgoing mesons
absorbed and emitted by the same quark (different quarks),
respectively.}\label{fig-1}
\end{figure}

\section{framework}\label{suc}

The tree diagrams calculated in the chiral quark model have been
shown in Fig. \ref{fig-1}. The reaction amplitude can be expressed
as the sum of the $s$-, $u$-, $t$-channel transition amplitudes:
\begin{eqnarray}
\mathcal{M}=\mathcal{M}_{s}+\mathcal{M}_{u}+\mathcal{M}_t \ .
\end{eqnarray}
The $s$- and $u$-channel transition amplitudes as shown in Fig.
\ref{fig-1} are given by
\begin{eqnarray}
\mathcal{M}_{s}=\sum_j\langle N_f |H^f_{m} |N_j\rangle\langle N_j
|\frac{1}{E_i+\omega_i-E_j}H^i_{m }|N_i\rangle, \label{sc}\\
\mathcal{M}_{u}=\sum_j\langle N_f |H^i_{m }
\frac{1}{E_i-\omega_f-E_j}|N_j\rangle\langle N_j | H^f_{m}
|N_i\rangle, \label{uc}
\end{eqnarray}
where $H^i_{m}$ and $H^f_{m}$ stand for the incoming and outgoing
meson-quark couplings, which might be described by the effective
chiral Lagrangian~\cite{Li:1997gda,qk3}
\begin{eqnarray}\label{coup}
H_m=
\frac{1}{f_m}\bar{\psi}_j\gamma^{j}_{\mu}\gamma^{j}_{5}\psi_j\vec{\tau}\cdot\partial^{\mu}\vec{\phi}_m,
\end{eqnarray}
where $\psi_j$ represents the $j$-th quark field in a hadron, and
$f_m$ is the meson's decay constant. The pseudoscalar-meson octet,
$\phi_m$, is written as
\begin{eqnarray}
\phi_m=\pmatrix{
 \frac{1}{\sqrt{2}}\pi^0+\frac{1}{\sqrt{6}}\eta & \pi^+ & K^+ \cr
 \pi^- & -\frac{1}{\sqrt{2}}\pi^0+\frac{1}{\sqrt{6}}\eta & K^0 \cr
 K^- & \bar{K}^0 & -\sqrt{\frac{2}{3}}\eta}.
\end{eqnarray}
In Eqs. (\ref{sc}) and (\ref{uc}), $\omega_i$ and $\omega_f$ are the
energies of the incoming and outgoing mesons, respectively.
$|N_i\rangle$, $|N_j\rangle$ and $|N_f\rangle$ stand for the
initial, intermediate and final states, respectively, and their
corresponding energies are $E_i$, $E_j$ and $E_f$, which are the
eigenvalues of the nonrelativistic Hamiltonian of constituent quark
model $\hat{H}$~\cite{Isgur78, Isgur:1977ky}.

The extracted transition amplitude for the $s$ channel
is~\cite{Zhong:2007fx,Zhong:2009}
\begin{eqnarray} \label{sac}
\mathcal{M}_{s}&&=\Big\{\sum_{n=0}\left(g_{s1}+\frac{1}{(-2)^{n}}
g_{s2}\right)\textbf{A}_{out}\cdot\textbf{A}_{in}
 \frac{F_s(n)}{n !}\mathcal{X}^n\nonumber
\\ && +\sum_{n=1}
\left(g_{s1}+\frac{1}{(-2)^{n}}g_{s2}\right) \frac{F_s(n)}{(n-1)
!}\mathcal{X}^{n-1}\nonumber\\
&&\times\left(\frac{\omega_{i}}{6\mu_q}\textbf{A}_{out}\cdot
\textbf{q}+\frac{\omega_{i}}{3m_q}\textbf{A}_{in}\cdot
\textbf{k}+\frac{\omega_{i}}{m_q}\frac{\omega_{i}}{2\mu_q}
\alpha^2\right)\nonumber\\&&+
\sum_{n=2}\left(g_{s1}+\frac{1}{(-2)^{n}}g_{s2}\right)\frac{\omega_{f}\omega_{i}}
{18m_q\mu_q}\textbf{k}\cdot\textbf{q}
\frac{F_s(n)}{(n-2)!}\mathcal{X}^{n-2}\ \ \ \nonumber\\
&&+\sum_{n=0}\left(g_{v1}+\frac{1}{(-2)^{n}}g_{v2}\right)i \vsig
\cdot(\textbf{A}_{out}\times\textbf{A}_{in}) \frac{F_s(n)}{n
!}\mathcal{X}^n\ \ \ \nonumber\\
&&-\sum_{n=2}\left(g_{v1}+\frac{1}{(-2)^{n}}g_{v2}\right)
\frac{\omega_{f}\omega_{i}}{18m_q\mu_q}i\vsig\cdot(\textbf{q}\times\textbf{k})
\ \ \nonumber\\
&& \times
\frac{F_s(n)}{(n-2)!}\mathcal{X}^{n-2}\Big\}e^{-(\textbf{k}^2+\textbf{q}^2)/6\alpha^2},
\end{eqnarray}
with
\begin{eqnarray}
\mathbf{A}_{in}=-\left( \omega_{i} \mathcal{K}_i+\frac{\omega_i}{6\mu_q}+1\right)\textbf{k},\\
\mathbf{A}_{out}=-\left(\omega_{f}\mathcal{K}_f+\frac{\omega_f}{6\mu_q}+1\right)\textbf{q},
\end{eqnarray}
where $\mathcal{K}_i\equiv1/(E_i+M_i)$,
$\mathcal{K}_f\equiv1/(E_f+M_f)$ and $m_q$ is the light quark mass.
The $\mathcal{X}$ is defined as $\mathcal{X}\equiv\textbf{k}\cdot
\textbf{q}/(3 \alpha^2)$, and the factor $F_s(n)$ is given by
expanding the energy propagator in Eq. (\ref{sac})  which leads to
\begin{eqnarray}
F_s(n)=\frac{M_n}{P_i \cdot k-nM_n \omega_h},
\end{eqnarray}
where $M_n$ is the mass of the intermediate baryons in the $n$-th
shell, while $\omega_h$ is the typical energy of the harmonic
oscillator; $P_i$ and $k$ are the four momenta of the initial
baryons and incoming mesons in the center-of-mass (c.m.) system,
respectively.

While the extracted transition amplitude for the $u$ channel is
~\cite{Zhong:2007fx,Zhong:2009}
\begin{eqnarray}\label{uac}
\mathcal{M}_{u}&=&-\Big\{\textbf{B}_{in}\cdot\textbf{B}_{out}\sum_{n=0}
\left[g^u_{s1}+(-2)^{-n}g^u_{s2}\right] \frac{F_u(n)}{n
!}\mathcal{X}^n \nonumber\\
&&+\left(\frac{\omega_{f}}{3m_q}\textbf{B}_{in}\cdot
\textbf{k}+\frac{\omega_{i}}{6\mu_q}\textbf{B}_{out}\cdot
\textbf{q}+\frac{\omega_{i}}{2\mu_q}\frac{\omega_{f}}{m_q}
\alpha^2\right)\nonumber\\
&&\times\sum_{n=1}\left[g^u_{s1}+(-2)^{-n}g^u_{s2}\right]
\frac{F_u(n)}{(n-1) !}\mathcal{X}^{n-1}\nonumber\\
&&+ \frac{\omega_{f}\omega_{i}}{18m_q\mu_q}\textbf{k}\cdot\textbf{q}
\sum_{n=2}\frac{F_u(n)}{(n-2)!}\left[g^u_{s1}+(-2)^{-n}g^u_{s2}\right]
\mathcal{X}^{n-2} \nonumber\\
&& + i\vsig
\cdot(\textbf{B}_{in}\times\textbf{B}_{out})\sum_{n=0}\left[g^u_{v1}+(-2)^{-n}g^u_{v2}\right]
\frac{F_u(n)}{n !}\mathcal{X}^n\nonumber\\
&&+\frac{\omega_{f}\omega_{i}}{18m_q\mu_q}i\vsig\cdot(\textbf{q}\times\textbf{k})
\sum_{n=2}\left[g^u_{v1}+(-2)^{-n}g^u_{v2}\right]\mathcal{X}^{n-2}\nonumber\\
&&\times\frac{F_u(n)}{(n-2)!} +i\vsig \cdot
\left[\frac{\omega_{f}}{3m_q}(\textbf{B}_{in}\times
\textbf{k})+\frac{\omega_{i}}{6\mu_q}(\textbf{q}\times\textbf{B}_{out}
)\right ]\ \ \ \ \ \nonumber\\
&&\sum_{n=1}\left[g^u_{v1}+(-2)^{-n}g^u_{v2}\right]
\mathcal{X}^{n-1} \frac{F_u(n)}{(n-1) !} \Big\}
e^{-(\textbf{k}^2+\textbf{q}^2)/6\alpha^2},
\end{eqnarray}
where, we have defined
\begin{eqnarray}
\mathbf{B}_{in}\equiv-\omega_i\left(K_f+K_j+\frac{1}{6\mu_q}\right)\mathbf{q}-(\omega_iK_j+1)\mathbf{k},\\
\mathbf{B}_{out}\equiv-\omega_f\left(K_i+K_j+\frac{1}{6\mu_q}\right)\mathbf{k}-(\omega_f
K_j+1)\mathbf{q},
\end{eqnarray}
with $\mathcal{K}_j\equiv1/(E_j+M_j)$. The factor $F_u(n)$ is given
by
\begin{eqnarray}
F_u(n)=\frac{M_n}{P_i \cdot q+nM_n \omega_h},
\end{eqnarray}
where $q$ stands for the four momenta of the outgoing mesons in the
c.m. system.

The $g$-factors appeared in the $s$- and $u$-channel amplitudes are
determined by~\cite{Zhong:2009}
\begin{eqnarray}
g_{s1}&\equiv&\langle N_f |\sum_{j} I^{f}_j I^{i}_j |N_i
\rangle/3,\\
g_{s2}&\equiv&\langle N_f |\sum_{i\neq j} I^{f}_i
I^{i}_j\vsig_i\cdot \vsig_j |N_i
\rangle/3,\\
g_{v1}&\equiv&\langle N_f |\sum_{j} I^{f}_i
I^{i}_j\vsig_{jz} |N_i\rangle/2,\\
g_{v2}&\equiv&\langle N_f |\sum_{i\neq j} I^{f}_i
I^{i}_j(\vsig_i\times \vsig_j)_z |N_i\rangle/2,\\
g^u_{s1}&\equiv&\langle N_f |\sum_{j}I^{i}_j I^{f}_j |N_i \rangle,\\
g^u_{s2}&\equiv&\langle N_f |\sum_{i\neq j} I^{i}_i
I^{f}_j\vsig_i\cdot \vsig_j |N_i
\rangle/3,\\
g^u_{v1}&\equiv&\langle N_f |\sum_{j}
I^{i}_j I^{f}_j \sigma^z_j |N_i\rangle,\\
g^u_{v2}&\equiv&\langle N_f |\sum_{i\neq j} I^{i}_i I^{f}_j
(\vsig_i\times \vsig_j)_z |N_i\rangle/2,
\end{eqnarray}
where $\vsig_j$ corresponds to the Pauli spin vector of the $j$-th
quark in a hadron, $I^i_j$ and $I^f_j$ are the isospin operators of
the initial and final mesons defined in~\cite{Zhong:2009}.

These $g$-factors can be derived in the SU(6)$\otimes$O(3) symmetry
limit. In Tab.~\ref{factor}, we have listed the $g$-factors for the
reactions $K^-p\rightarrow\Sigma^0\pi^0$, $\Lambda\pi^0$ and
$\bar{K}^0n$. From these factors, we can see some interesting
features of these reactions. For example, it is found that in the
reactions $K^-p\rightarrow \Sigma^0\pi^0$ and $\Lambda\pi^0$, the
$K^-$- and $\pi^0$-mesons can not couple to the same quark of a
$s$-channel intermediate state (i.e., $g_{s1}=g_{v1}=0$), which
leads to a strong suppression of the $s$-channel contributions.
However, for the $u$ channel the kaon and pion can couple to not
only the same quark but also different quarks of a baryon. Thus, the
$u$ channel could contribute a large background to these two
processes. While for the charge-exchange reaction $K^-p\rightarrow
\bar{K}^0 n$, there are no $u$-channel contributions (i.e.,
$g^u=0$), and only the $s$-channel amplitude $M_3^s$ survives for
the isospin selection rule (i.e., $g_{s2}=g_{v2}=0$).

In this work, we consider the vector-exchange and the
scalar-exchange for the $t$-channel backgrounds. The vector
meson-quark and scalar meson-quark couplings are given by
\begin{eqnarray}\label{coup}
H_V&=& \bar{\psi}_j\left(a\gamma^{\nu}+\frac{
b\sigma^{\nu\lambda}\partial_{\lambda}}{2m_q}\right)V_{\nu} \psi_j,\\
H_S&=&g_{Sqq}\bar{\psi}_j\psi_jS,
\end{eqnarray}
where $V$ and $S$ stands for the vector and scalar fields,
respectively. The constants $a$, $b$ and $g_{Sqq}$ are the vector,
tensor and scalar coupling constants, respectively. They are treated
as free parameters in this work.

On the other hand, the $VPP$ and $SPP$ couplings ($P$ stands for a
pseudoscalar-meson) are adopted as \cite{PPV,Wu:2007fc}
\begin{eqnarray}
H_{VPP}&=&-iG_VTr([\phi_m,\partial_\mu\phi_m]V^{\mu}),\\
H_{SPP}&=&\frac{g_{SPP}}{2m_\pi}\partial_\mu\phi_m\partial^\mu
\phi_m S,
\end{eqnarray}
where $G_V$ is the coupling constant to be determined by
experimental data.

For the case of the vector meson exchange, the $t$-channel amplitude
in the quark model is given by
\begin{eqnarray}
\mathcal{M}^V_t=\mathcal{O}^t_V\frac{1}{t-M_{V}^{2}}e^{-(\mathbf{q}-\mathbf{k})^2/(6\alpha^2)},\label{t-v}
\end{eqnarray}
where $e^{-(\mathbf{q}-\mathbf{k})^2/(6\alpha^2)}$ is a form factor
deduced from the quark model, and $M_{V}$ is the vector-meson mass.
The amplitude $\mathcal{O}^t_V$ is given by
\begin{eqnarray}
\mathcal{O}^t_V&=&G_v a[g^s_{t}(\mathcal{H}_0+\mathcal{H}_1
\mathbf{q}\cdot \mathbf{k})+g^v_t\mathcal{H}_2
i\vsig\cdot(\mathbf{q}\times \mathbf{k})]\nonumber\\
&& +\mathrm{tensor\ \ term},\label{tt-v}
\end{eqnarray}
where we have defined
\begin{eqnarray}
\mathcal{H}_0&\equiv&E_0\left(1+\frac{\mathcal{K}_f}{6\mu_q}\mathbf{q}^2-\frac{\mathcal{K}_i}{6\mu_q}\mathbf{k}^2
+\frac{1}{4\mu_q^2}\left[\frac{\alpha^2}{3}+\frac{1}{9}(\mathbf{q}^2+\mathbf{k}^2)\right]\right)\nonumber\\
&&+\frac{1}{3\mu_q}(\mathbf{q}^2-\mathbf{k}^2)+\mathcal{K}_f\mathbf{q}^2+\mathcal{K}_i\mathbf{k}^2,\\
\mathcal{H}_1&\equiv&E_0\left[\mathcal{K}_i\mathcal{K}_f-\frac{\mathcal{K}_f}{6\mu_q}+\frac{\mathcal{K}_i}{6\mu_q}
-\frac{1}{2\mu_q^2}\right]+\mathcal{K}_f+\mathcal{K}_i,\\
\mathcal{H}_2&\equiv&E_0\left[\mathcal{K}_f\mathcal{K}_i-\frac{\mathcal{K}_f}{6\mu_q}
+\frac{\mathcal{K}_i}{6\mu_q}\right]+\mathcal{K}_i +\mathcal{K}_f,
\end{eqnarray}
with $E_0=\omega_i+\omega_f$. The tensor term of the $t$-channel
vector-exchange amplitude is less important than that of vector
term. In the calculations, we find the results are insensitive to
the tensor term, thus, its contributions are neglected for
simplicity. In Eq.(\ref{tt-v}), we have defined $g^s_t\equiv \langle
N_f|\sum^3_{j=1}I^{ex}_j|N_i\rangle$, and $g^v_t\equiv \langle
N_f|\sum^3_{j=1}\sigma_j I^{ex}_j|N_i\rangle$, which can be deduced
from the quark model, where, $I^{ex}_j$ is the isospin operator of
exchanged meson. For the $K^-p\rightarrow\Sigma^0 \pi^0,\Lambda
\pi^0$ processes, the vector $K^{*+}$-exchange is considered, and
for the $K^-p\rightarrow n \bar{K}^0$ process, the vector
$\rho^+$-exchanged is considered.

While, for the case of the scalar meson exchange, the  $t$-channel
amplitude in the quark model is written as
\begin{eqnarray}
\mathcal{M}^S_t=\mathcal{O}^t_S\frac{1}{t-m^2_{S}
}e^{-(\mathbf{q}-\mathbf{k})^2/(6\alpha^2)},
\end{eqnarray}
where $m_{S}$ is the scalar-meson mass, and the $\mathcal{O}^t_S$ is
given by
\begin{eqnarray}
\mathcal{O}^t_S&\simeq &\frac{g_{S PP} g_{S
qq}}{2m_\pi}(\omega_i\omega_f-\mathbf{q}\cdot
\mathbf{k})[g^s_{t}(\mathcal{A}_0+\mathcal{A}_1\mathbf{q}\cdot
\mathbf{k})\nonumber\\&& +g^v_t \mathcal{A}_2
i\vsig\cdot(\mathbf{q}\times \mathbf{k}) ]\label{tt-s},
\end{eqnarray}
with
\begin{eqnarray}
\mathcal{A}_0&\equiv&1-\frac{\mathcal{K}_f}{6\mu_q}\mathbf{q}^2+\frac{\mathcal{K}_i}{6\mu_q}\mathbf{k}^2
-\frac{1}{4\mu_q^2}\left[\frac{\alpha^2}{3}+\frac{\mathbf{q}^2+\mathbf{k}^2}{9}\right],\\
\mathcal{A}_1&\equiv&
-\mathcal{K}_i\mathcal{K}_f+\frac{1}{6\mu_q}\mathcal{K}_f-\frac{1}{6\mu_q}\mathcal{K}_i+\frac{1}{18\mu_q^2},\\
\mathcal{A}_2&\equiv&-\mathcal{K}_i\mathcal{K}_f+\frac{1}{6\mu_q}\mathcal{K}_f-\frac{1}{6\mu_q}\mathcal{K}_i.
\end{eqnarray}
In Eq.(\ref{tt-s}), we have neglected the higher order terms. In
this work, the scalar $\kappa$-exchange is considered for the
$K^-p\rightarrow\Sigma^0 \pi^0,\Lambda \pi^0$ processes, while the
scalar $a_0(980)$-exchange is considered for the $K^-p\rightarrow n
\bar{K}^0$ process.

\section{separation of the resonance contributions}\label{sres}

It should be remarked that the amplitudes in terms of the harmonic
oscillator principal quantum number $n$ are the sum of a set of
SU(6) multiplets with the same $n$. To see the contributions of an
individual resonance listed in Tab.~\ref{Resonance}, we need to
separate out the single-resonance-excitation amplitudes within each
principal number $n$ in the $s$ channel.

We have noticed that the transition amplitude has a unified
form~\cite{Hamilton:1963zz}:
\begin{eqnarray}
\mathcal{O}=f(\theta)+ig(\theta) \vsig \cdot \textbf{n},
\end{eqnarray}
where $\textbf{n}\equiv\textbf{q}\times
\textbf{k}/|\textbf{k}\times\textbf{q}|$. The non-spin-flip   and
spin-flip amplitudes $f(\theta)$ and $g(\theta)$ can be expanded in
terms of the familiar partial wave amplitudes $T_{l\pm }$ for the
states with $J=l\pm 1/2$:
\begin{eqnarray}
f(\theta)&=&\sum^{\infty}_{l=0} [(l+1)T_{l+}+lT_{l-}]P_l(\cos\theta),\label{P1} \\
g(\theta)&=&\sum^{\infty}_{l=0} [T_{l-}-T_{l+}]\sin\theta
P'_l(\cos\theta).\label{P2}
\end{eqnarray}

Combining Eqs.~(\ref{P1}) and~(\ref{P2}), firstly, we can separate
out the partial waves with different $l$ in the same $n$. For
example, in the $n=0$ shell, only the $P$ ($l=1$) wave contributes
to the reaction; in the $n=1$ shell, both $S$ ($l=0$) and $D$
($l=2$) waves contribute to the reaction; and in the $n=2$ shell,
only the $P$ and $F$ waves are involved in the process. The
separated partial amplitudes, $\mathcal{O}_n(l)$, up to the $n=2$
shell are given by~\cite{Zhong:2009,Zhong:2007fx}
\begin{eqnarray}
\mathcal{O}_0(P)&=&(g_{s1}+g_{s2})\textbf{A}_{out}\cdot\textbf{A}_{in}\nonumber\\
&& +(g_{v1}+g_{v2})i\vsig \cdot(\textbf{A}_{out}\times\textbf{A}_{in}),\\
\mathcal{O}_1(S)&=&\left(g_{s1}-\frac{1}{2}g_{s2}\right)\Big(|\mathbf{A}_{out}|\cdot|\mathbf{A}_{in}
|\frac{|\mathbf{k}||\mathbf{q}|}{9\alpha^2}
+\frac{\omega_i}{6\mu_q}\mathbf{A}_{out}\cdot
\mathbf{q}\nonumber\\
&&+\frac{\omega_f}{6\mu_q}\mathbf{A}_{in}\cdot
\mathbf{k}+\frac{\omega_i\omega_f}{4\mu_q\mu_q}\alpha^2\Big),\\
\mathcal{O}_1(D)&=&\left(g_{s1}-\frac{1}{2}g_{s2}\right)|\mathbf{A}_{out}|\cdot|\mathbf{A}_{in}|(3\cos^2\theta-1)
\frac{|\mathbf{k}||\mathbf{q}|}{9\alpha^2}\nonumber\\
&&+\left(g_{v1}-\frac{1}{2}g_{v2}\right)i\vsig
\cdot(\textbf{A}_{out}\times\textbf{A}_{in})\frac{\mathbf{k}\cdot\mathbf{q}}{3\alpha^2},\\
\mathcal{O}_2(P)&=&\left(g_{s1}+\frac{1}{4}g_{s2}\right)
\Big(|\textbf{A}_{out}||\textbf{A}_{in}|
\frac{|\textbf{k}||\textbf{q}|}{10
\alpha^2}+\frac{\omega_{i}}{6\mu_q}\textbf{A}_{out}\cdot
\textbf{q}\nonumber\\
&&+\frac{\omega_{f}}{6\mu_q}\textbf{A}_{in}\cdot
\textbf{k}+\frac{\omega_{f}}{\mu_q}\frac{\omega_{i}}{\mu_q}
\frac{\alpha^2}{3}\Big) \frac{|\textbf{k}||\textbf{q}|}{3
\alpha^2}\cos
\theta\nonumber\\
&&-\left(g_{v1}+\frac{1}{4}g_{v2}\right)
\frac{\omega_{f}\omega_{i}}{(6\mu_q)^2}i\vsig\cdot(\textbf{q}\times\textbf{k})\nonumber\\
&&+\frac{1}{10}\left(g_{v1}+\frac{1}{4}g_{v2}\right) i \vsig
\cdot(\textbf{A}_{out}\times\textbf{A}_{in})\left(\frac{|\textbf{k}||\textbf{q}|}{3 \alpha^2}\right)^2,\\
\mathcal{O}_2(F)&=&\left(g_{s1}+\frac{1}{4}g_{s2}\right)\frac{1}{2}|\textbf{A}_{out}||\textbf{A}_{in}|
\left(\cos^3 \theta-\frac{3}{5}\cos
\theta\right)\nonumber\\
&& \times\left(\frac{|\textbf{k}||\textbf{q}|}{3
\alpha^2}\right)^2+\left(g_{v1}+\frac{1}{4}g_{v2}\right)i \vsig
\cdot(\textbf{A}_{out}\times\textbf{A}_{in}) \nonumber\\
&& \times\frac{1}{2}\left(\cos^2 \theta -\frac{1}{5}
\right)\left(\frac{|\textbf{k}||\textbf{q}|}{3 \alpha^2}\right)^2.
\end{eqnarray}

Then, using the Eqs. (\ref{P1}) and (\ref{P2}) again, we can
separate out the partial amplitudes $\mathcal{O}_n(l)$ for the
states with different $J^P$ in the same $l$ as well. For example, we
can separate out the resonance amplitudes with $J^p=3/2^-$ [i.e.,
$\mathcal{O}_1(D_{I3})$] and $J^p=5/2^-$ [i.e.,
$\mathcal{O}_1(D_{I5})$] from the amplitude $\mathcal{O}_1(D)$.

Finally, we should sperate out the partial amplitudes with the same
quantum numbers $n$, $l$, $J^P$ in the different representations of
the constituent quark model. We notice that the resonance transition
strengths in the spin-flavor space are determined by the matrix
element $\langle N_f|H^f_m |N_j\rangle\langle N_j|H^i_m|N_i\rangle$.
Their relative strengths $g_R$ ($R\equiv
l_{I2J}[N_6,^{2\mathbf{S}+1}N_3,\textbf{L}]$) can be explicitly
determined by the following relation:
\begin{eqnarray}
\frac{g_{l_{I2J}[N_6,^{2\mathbf{S}+1}N_3,L]}}{g_{l_{I2J}[N'_6,^{2\mathbf{S}'+1}N'_3,L']}}&=&\frac{\langle
N_f|I^{f}_3\sigma_{3z}|l_{I2J}[N_6,^{2\mathbf{S}+1}N_3,L]\rangle}{\langle
N_f|I^{f}_3\sigma_{3z}|l_{I2J}[N'_6,^{2\mathbf{S}'+1}N'_3,L']\rangle}\nonumber\\
&&\cdot\frac{\langle
l_{I2J}[N_6,^{2\mathbf{S}+1}N_3,L]|I^i_3\sigma_{3z}|N_i\rangle}{\langle
l_{I2J}[N'_6,^{2\mathbf{S}'+1}N'_3,L']|I^i_3\sigma_{3z}|N_i
\rangle},\label{abcd}
\end{eqnarray}
At last, we obtain the single-resonance-excitation amplitudes
$\mathcal{O}_R$ by the relation:
\begin{eqnarray}
\mathcal{O}(n,l,J)=\sum_R\mathcal{O}_R(n,l,J)=\sum_Rg_R\mathcal{O}(n,l,J).
\end{eqnarray}
In this work, the values of $g_R$ for the reactions
$K^-p\rightarrow\Sigma^0\pi^0$, $\Lambda\pi^0$ and $\bar{K}^0n$ have
been derived in the symmetric quark model, which have been listed in
Tab.~\ref{factor}.

\begin{table}[ht]
\caption{Various g and $g_R$ factors extracted in the symmetric
quark model. } \label{factor}
\begin{tabular}{|c|c|c|c|c|c|c|c|c|c|c|c }\hline
  \multicolumn{2}{|c|}{$K^-p\rightarrow\Sigma^0 \pi^0$}
 & \multicolumn{2}{|c|}{$K^-p\rightarrow \Lambda \pi^0$}
 & \multicolumn{2}{|c|}{$K^-p\rightarrow \bar{K}^0 n$}   \\ \hline
 factor       &  value         &   factor                 &  value                      &  factor                 &  value               \\
 \hline
 $g^u_{s1}$              &    $1/2$                           &   $g^u_{s1}$             & $\sqrt{3}/2$                     &  $g_{s1}$               &1                                \\
 $g^u_{s2}$              &     1                             &   $g^u_{s2}$             & $\sqrt{3}/3$                    & $g_{v1}$                & 5/3                              \\
 $g^u_{v1}$              &     $-1/6$                             &   $g^u_{v1}$             & $\sqrt{3}/2$                   & $g_{S_{01}[70,^21]}$    &  27/36         \\
 $g^u_{v2}$              &      $-1$                             &   $g^u_{v2}$             & $\sqrt{3}/3$                     & $g_{S_{01}[70,^28]}$    &   27/36       \\
 $g_{s2}$                &      1                            &   $g_{s2}$               & $\sqrt{3}/3$                 & $g_{S_{11}[70,^28]}$    & $-1/36$                            \\
 $g_{v2}$                &       1                              &   $g_{v2}$               & $-\sqrt{3}/3$                 & $g_{S_{11}[70,^48]}$    &  $-16/36$                      \\

 $g_{S_{01}[70,^21]}$    &    3/2                       &   $g_{S_{11}[70,^28]}$   & $-1/6$                                   &  $g_{S_{11}[70,^210]}$  &  $-1/36$         \\
 $g_{S_{01}[70,^28]}$    &    $-1/2$                         &   $g_{S_{11}[70,^48]}$   & $4/6$                                     &  $g_{D_{03}[70,^21]}$   & 135/252         \\
 $g_{D_{03}[70,^21]}$    &     3/2                          &   $g_{S_{11}[70,^210]}$   & $-1/6$                                   &  $g_{D_{03}[70,^28]}$   &  135/252                        \\
 $g_{D_{03}[70,^28]}$    &     $-1/2$                         &   $g_{D_{13}[70,^28]}$   & $5/6$                                     &  $g_{D_{13}[70,^28]}$   &   $-5/252$                       \\
                         &                                   &   $g_{D_{13}[70,^48]}$   & $-4/6$                                    &  $g_{D_{13}[70,^48]}$   &  $-8/252$                         \\
                          &                                  &   $g_{D_{13}[70,^210]}$  & $-5/6$                                    &  $g_{D_{13}[70,^210]}$  & $-5/252$                          \\

                         &                                &                         &                                           &  $g_{\Lambda}$          &  27/26                          \\
                         &                               &                         &                                           &  $g_{\Sigma}$           &  $-1/26$                          \\
\hline
\end{tabular}
\end{table}

Taking into account the width effects of the resonances, the
resonance transition amplitudes of the $s$ channel can be generally
expressed as \cite{qk3,Zhong:2007fx}
\begin{eqnarray}
\mathcal{M}^s_R=\frac{2M_R}{s-M^2_R+iM_R
\Gamma_R(\textbf{q})}\mathcal{O}_Re^{-(\textbf{k}^2+\textbf{q}^2)/6\alpha^2},
\label{stt}
\end{eqnarray}
where $\Gamma_R(\textbf{q})$ is an energy-dependent width introduced
for the resonances in order to take into account the off-mass-shell
effects in the reaction. It is adopted as \cite{qkk,qk3,qk4}
\begin{eqnarray}
\Gamma_R(\textbf{q})=\Gamma_R\frac{\sqrt{s}}{M_R}\sum_i x_i
\left(\frac{|\textbf{q}_i|}{|\textbf{q}^R_i|}\right)^{2l+1}
\frac{D(\textbf{q}_i)}{D(\textbf{q}^R_i)},
\end{eqnarray}
where $|\textbf{q}^R_i|=((M_R^2-M_b^2+m_i^2)/4M_R^2-m_i^2)^{1/2}$,
and $|\textbf{q}_i|=((s-M_b^2+m_i^2)/4s-m_i^2)^{1/2}$; $x_i$ is the
branching ratio of the resonance decaying into a meson with mass
$m_i$ and a baryon with mass $M_b$, and $\Gamma_R$ is the total
decay width of the resonance with mass $M_R$.
$D(\textbf{q})=e^{-\textbf{q}^2/3\alpha^2}$ is a fission barrier
function.

\section{CALCULATION AND ANALYSIS} \label{MP}

\subsection{Parameters}

\begin{table}[ht]
\caption{The strength parameters $C_R$ determined by the
experimental data.} \label{param C}
\begin{tabular}{|c|c|c|c|c|c| }\hline\hline
Parameter    &\ \  $K^-p \rightarrow \Sigma^0\pi^0$ \ \ & \ \ $K^-
p\rightarrow \Lambda\pi^0$ &
$K^- p\rightarrow \bar{K}^0n$ \\
\hline
$C^{[70,^21]}_{S_{01}(1405)}$& 1.13$^{+0.17}_{-0.05}$     &...    & 0.72$^{+0.03}_{-0.06}$     \\
$C^{[70,^28]}_{S_{01}(1670)}$& 0.33$^{+0.05}_{-0.08}$     &...     & 0.08$^{+0.02}_{-0.03}$    \\
$C^{[70,^28]}_{S_{11}(1630)}$& ...      &1.00     & 1.00       \\
$C^{[70,^48]}_{S_{11}(1750)}$& ...      &0.86$^{+0.08}_{-0.12}$    & 0.50$^{+0.15}_{-0.15}$     \\
$C^{[70,^210]}_{S_{11}(1810)}$& ...     &1.00     & 1.00     \\
$C^{[70,^21]}_{D_{03}(1520)}$& 2.49$^{+0.06}_{-0.05}$     &...    & 2.87$^{+0.15}_{-0.15}$    \\
$C^{[70,^28]}_{D_{03}(1690)}$& 1.00     &...    & 0.30$^{+0.08}_{-0.04}$     \\
$C^{[70,^28]}_{D_{13}(1670)}$& ...     &1.00    & 5.00$^{+1.00}_{-2.00}$     \\
$C^{[70,^48]}_{D_{13}(1740)}$& ...     &1.00     & 1.00  \\
$C^{[70,^210]}_{D_{13}(1780)}$& ...     &1.00    & 1.00    \\
$C^{[70,^48]}_{D_{15}(1775)}$& ...     &0.78$^{+0.20}_{-0.13}$     & 1.00      \\
$C_{u}$                     & 0.68$^{+0.05}_{-0.08}$     &0.95$^{+0.03}_{-0.02}$     & ...     \\
$\sqrt{\delta_{m_i}\delta_{m_f}}$& 0.99$\pm$ 0.01     &1.13$\pm$ 0.01     & 1.08$\pm$ 0.01      \\
\hline
\end{tabular}
\end{table}

\begin{table}[ht]
\caption{Breit-Wigner masses $M_R$ (MeV) and widths $\Gamma_R$ (
MeV) for the resonances. } \label{BW}
\begin{tabular}{|c|c|c|c|c|c|c|c| }\hline\hline
$[N_6,^{2S+1}N_3,n, \mathbf{L}]$ &\ \  $l_{I,2J}$ \ \ & $M_R$ & \ \
$\Gamma_R$ \ \
&\ \ $M_R$ (PDG)\ \ & \ \ $\Gamma_R$ (PDG)\ \  \\
\hline
$[\textbf{70},^2 1,  1, \mathbf{1}]$& $S_{01}$    & 1410  & 80  &$1406\pm 4 $   &$50\pm 2$   \\
 & $D_{03}$    & 1519  & 19  &$1520\pm 1 $   &$16\pm 1$\\
\hline
$[\textbf{70},^2 10, 1,\mathbf{1}]$& $S_{11}$    & 1810  & 200 & ...    & ...  \\
 & $D_{13}$    & 1780  & 150 &      ...           &  ...             \\
\hline
$[\textbf{70},^28 , 1,\mathbf{1}]$& $S_{01}$    & 1674   & 50  & $1670\pm 10$   &$25\sim50$ \\
 & $D_{03}$    & 1685   & 62  & $1690\pm 5$    &$60\pm 10$ \\
 & $S_{11}$    & 1631   & 102  & $1620$          &   $10\sim 110$ \\
 & $D_{13}$    & 1674   & 52  &   $1675\pm 10$   &$60\pm 20$ \\
\hline
$[\textbf{70},^48 , 1,\mathbf{1}]$& $S_{11}$    & 1770  & 90 & $1765\pm 35$    & $60\sim 160$ \\
 & $D_{13}$    & 1740  &  80 & ...          & ... \\
 & $D_{15}$    & 1775  & 105 & $1775\pm 5$    &$120\pm 15$ \\\hline

$[\textbf{56},^28 , 2,\mathbf{0}]$& $P_{01}$    & 1600  & 150 &  $1630\pm 70$    &$150\pm 100$ \\
                                & $P_{11}$    & 1660    & 160  & $1660\pm 30$    &$40\sim 200$ \\
\hline
\end{tabular}
\end{table}

With the transition amplitudes derived from the previous section,
the differential cross section and polarization of final baryon can
be calculated by
\begin{eqnarray}
\frac{d\vsig}{d\Omega}&=&\frac{(E_i+M_i)(E_f+M_f)}{64\pi^2 s
(2M_i)(2M_f)}\frac{|\textbf{q}|}{|\textbf{k}|}\frac{M_N^2}{2}\nonumber\\
&&\times\sum_{\lambda_i,\lambda_f}\left|\left[
\frac{\delta_{m_i}}{f_{m_i}}\frac{\delta_{m_f}}{f_{m_f}}(\mathcal{M}_s+\mathcal{M}_u)+
\mathcal{M}_t\right]_{\lambda_f,\lambda_i}\right|^2 ,\\
P&=&2\frac{\mathrm{Im}[f(\theta)g^*(\theta)]}{|f(\theta)|^2+|g(\theta)|^2},
\end{eqnarray}
where $\lambda_i=\pm 1/2$ and $\lambda_f=\pm 1/2$ are the helicities
of the initial and final state baryons, respectively.
$\delta_{m_i}\delta_{m_f}$ is a global parameter accounting for the
flavor symmetry breaking effects arising from the $quark-meson$
couplings, which is to be determined by experimental data. $f_{m_i}$
and $f_{m_f}$ are the initial and final meson decay constants,
respectively.

In the calculation, the universal value of harmonic oscillator
parameter $\alpha=0.4$ GeV is adopted. The masses of the $u$, $d$,
and $s$ constituent quarks are set as $m_u=m_d=330$ MeV, and
$m_s=450$ MeV, respectively. The decay constants for $\pi$, and $K$
are adopted as $f_\pi=132$ MeV and $f_K=160$ MeV, respectively.

In our framework, the resonance transition amplitude,
$\mathcal{O}_R$, is derived in the SU(6)$\otimes$O(3) symmetric
quark model limit. In reality, the SU(6)$\otimes$O(3) symmetry is
generally broken due to e.g. spin-dependent forces in the
quark-quark interaction. As a consequence, configuration mixings
would occur, and an analytic solution cannot be achieved. To take
into account the breaking of that symmetry, an empirical
way~\cite{He:2008ty,Saghai:2001yd} is to introduce a set of coupling
strength parameters, $C_R$, for each resonance amplitude,
\begin{eqnarray}
\mathcal{O}_R\rightarrow C_{R} \mathcal{O}_{R} \ ,
\end{eqnarray}
where $C_R$ should be determined by fitting the experimental
observables. In the SU(6)$\otimes$O(3) symmetry limit one finds
$C_R=1$, while deviations of $C_R$ from unity imply the
SU(6)$\otimes$O(3) symmetry breaking. The determined values of $C_R$
have been listed in Tab.~\ref{param C}. For the uncertainties of the
data, the parameters listed in Tab.~\ref{param C} have some
uncertainties as well. To know some uncertainties of a main
parameter, we vary it around its central value until the predictions
are inconsistent with the data within their uncertainties. The
obtained uncertainties for the main parameters have been given in
Tab.~\ref{param C} as well.

In the $t$ channel, two coupling constants, $G_Va$ from
vector-exchange and $g_{SPP}g_{Sqq}$ from scalar-exchange, are
considered as free parameters. By fitting the data, we found that
$G_Va\simeq4$ and $g_{SPP}g_{Sqq}\simeq117$ for the
$K^-p\rightarrow\Sigma^0 \pi^0,\Lambda \pi^0,n \bar{K}^0$ processes.

In the calculations, the $n=2$ shell $S=-1$ resonances in the $s$
channel are treated as degeneration, their degenerate mass and width
are taken as $M=1800$ MeV and $\Gamma=100$ MeV, since in the low
energy region the contributions from the $n=2$ shell are not
significant. In the $u$ channel, the intermediate states are the
nucleon and its resonances. It is found that contributions from the
$n\geq 1$ shells of the $u$ channel are negligibly small, thus, the
masses of the intermediate states for these shells are also treated
as degeneration. In this work, we take $M_1=1650$ MeV ($M_2=1750$
MeV) for the degenerate mass of the $n=1$ ($n=2$) shell nucleon
resonances. By fitting the data, we obtain the masses and widths of
the main strange resonances in the $s$ channel, which are listed in
Tab. \ref{BW}. Our results show that the resonance parameters are in
agreement with the PDG values.

\begin{widetext}
\begin{center}
\begin{figure}[ht]
\centering \epsfxsize=15 cm \epsfbox{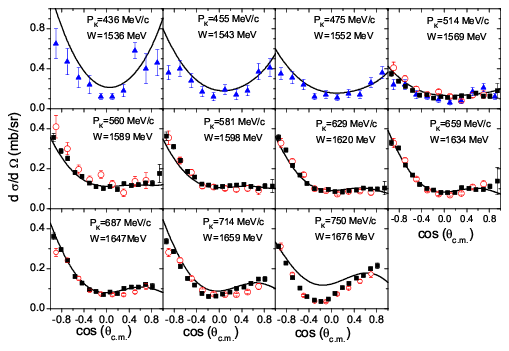} \epsfxsize=15 cm
\epsfbox{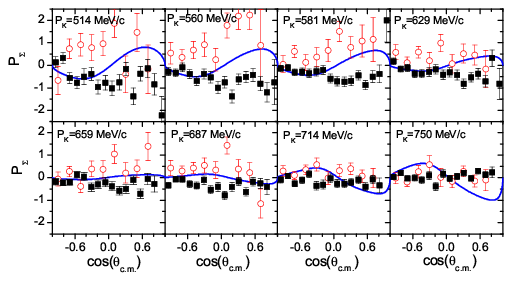} \caption{ (Color online) Differential cross
sections (upper panel) and $\Sigma^0$ polarizations (lower panel) of
the $K^-p\rightarrow \Sigma^0 \pi^0$ process compared with the data
are from \cite{Manweiler:2008zz} (open circles), \cite{Prakhov:2008}
(squares) and \cite{armen:1970zh} (triangles).}\label{sigmapi}
\end{figure}
\end{center}
\end{widetext}

\subsection{$K^-p\rightarrow\Sigma^0\pi^0$}

\begin{figure}[ht]
\centering \epsfxsize=8.0 cm \epsfbox{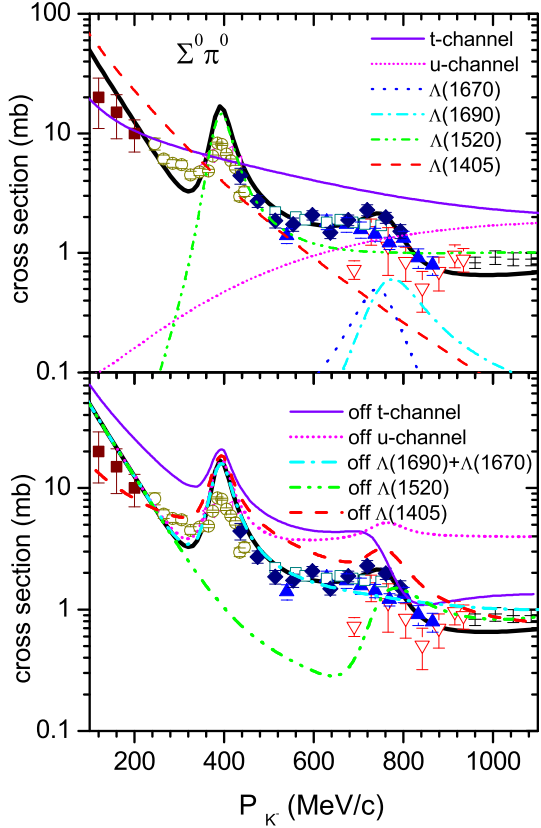} \caption{(Color
online) Cross section of the $K^-p\rightarrow \Sigma^0 \pi^0$
process. The bold solid curves are for the full model calculations.
Data are from Refs. \cite{Baxter:1974zs} (down-triangles),
\cite{Mast:1974sx} ( solid circles), \cite{armen:1970zh} (solid
diamonds), \cite{Berley:1996zh} (left-triangles),
\cite{London:1975av} (up-triangles), \cite{Manweiler:2008zz} (open
squares), and \cite{Kim:1965} (solid squares). In the upper panel,
exclusive cross sections for $\Lambda(1405)S_{01}$,
$\Lambda(1520)D_{03}$, $\Lambda(1670)S_{01}$, $\Lambda(1690)D_{03}$,
$t$ channel, and $u$ channel are indicated explicitly by the legends
in the figures. In the lower panel, the results by switching off the
contributions of $\Lambda(1405)S_{01}$, $\Lambda(1520)D_{03}$,
$\Lambda(1670)S_{01}$, $\Lambda(1690)D_{03}$, $t$ channel, and $u$
channel are indicated explicitly by the legends in the
figures.}\label{csa}
\end{figure}

The $K^-p\rightarrow\Sigma^0\pi^0$ process provides us a rather
clear channel to study the $\Lambda$ resonances, because only the
$\Lambda$ resonances contribute here for the isospin selection rule.
The low-lying $\Lambda$ resonances classified in the quark model are
listed in Tab. \ref{Resonance}, from which we see that in $n=0$
shell, only the $\Lambda$ pole contributes to the process. In the
$n=1$ shell, two $S$ wave states (i.e.,
$[70,^28]\Lambda(1670)S_{01}$, $[70,^21]\Lambda(1405)S_{01}$), and
two $D$ wave states (i.e., $[70,^21]\Lambda(1520)D_{03}$,
$[70,^28]\Lambda(1690)D_{03}$) contribute to the reaction. The
excitations of $[70,^48]$ are forbidden for the $\Lambda$-selection
rule \cite{Zhao:2006an,Isgur:1978xb,Hey:1974nc}. In our previous
work~\cite{Zhong:2009}, we have studied the $K^-p\rightarrow\Sigma^0
\pi^0$ process. For more accurate data of $K^-p\rightarrow \Sigma^0
\pi^0,\bar{K}^0n$ are reported recently, which can be used to
further constrain the properties of the $\Lambda$ resonances. In
this work we revisit the $K^-p\rightarrow\Sigma^0 \pi^0$ process.
The differential cross sections, $\Sigma^0$ polarizations and total
cross sections compared with the data are shown in
Figs.~\ref{sigmapi} and ~\ref{csa}. They are well described with the
parameters determined by fitting the 112 data of the differential
cross sections from Ref.~\cite{Prakhov:2008}. The $\chi^2$ per datum
point is about $\chi^2/N=3.4$. The main conclusions of our previous
study~\cite{Zhong:2009} still hold as compared with those of the
present work. In this work, the descriptions of the differential
cross sections at forward angles are obviously improved (see
Fig.~\ref{sigmapi}). From Fig.~\ref{sigmapi}, it is seen that the
measurements of the $\Sigma^0$ polarizations
from~\cite{Manweiler:2008zz} and \cite{Prakhov:2008} are not
consistent with each other. Our theoretical results of $\Sigma^0$
polarizations show some agreement with the measurements
from~\cite{Prakhov:2008} at backward angles.

\begin{figure}[ht]
\centering \epsfxsize=8.0 cm \epsfbox{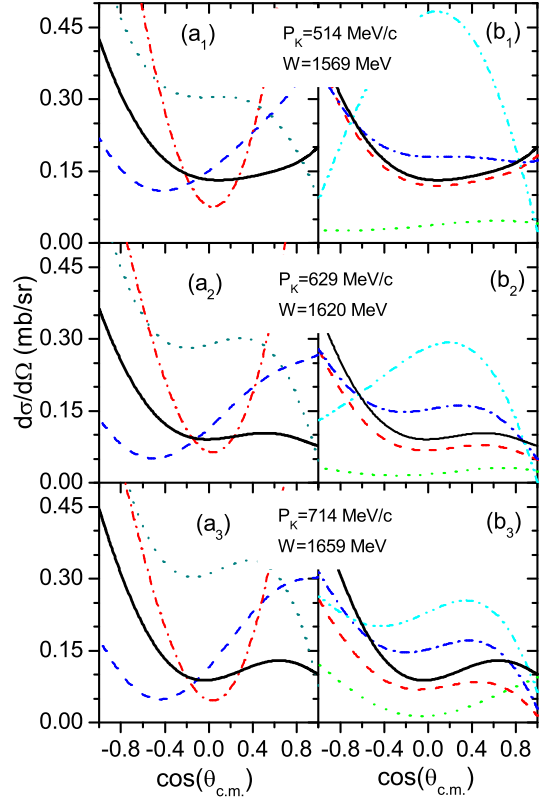} \caption{(Color
online) Effects of backgrounds and individual resonances on the
differential cross sections at three energies for the
$K^-p\rightarrow \Sigma^0 \pi^0$ process. The bold solid curves are
for the full model calculations. In panels ($a_1$)-($a_3$), the
dashed, dotted, and dash-dotted are for the results given by
switching off the contributions from the $\Lambda$ pole, $u$ channel
and $t$ channel, respectively. In panels ($b_1$)-($b_3$), the
dotted, dashed, dash-dotted, dash-dot-doted curves stand for the
results given by switching off the contributions from
$\Lambda(1520)D_{03}$, $\Lambda(1690)D_{03}$, $\Lambda(1670)S_{01}$
and $\Lambda(1405)S_{01}$, respectively.}\label{spi}
\end{figure}

\begin{figure}[ht]
\centering \epsfxsize=8.0 cm \epsfbox{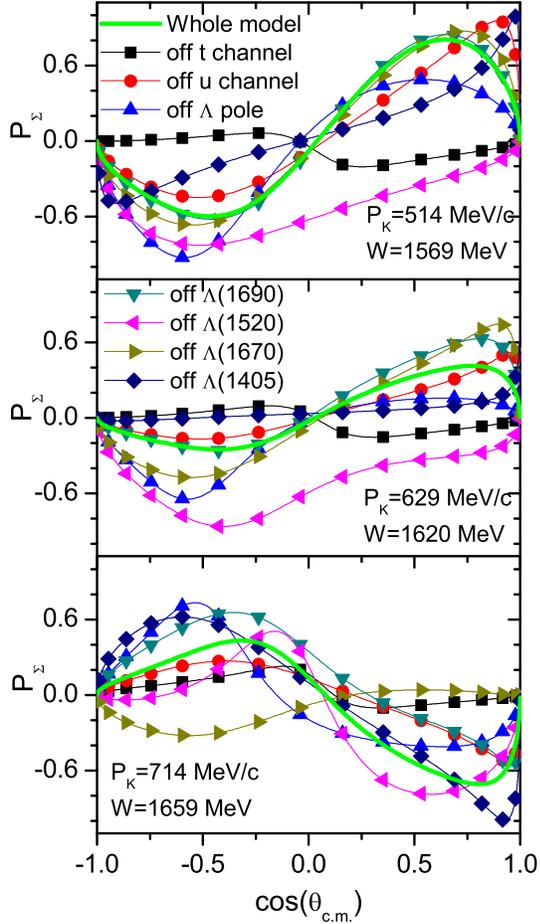} \caption{(Color
online) Effects of backgrounds and individual resonances on the
$\Sigma^0$ polarizations at three energies for the $K^-p\rightarrow
\Sigma^0 \pi^0$ process. The bold solid curves are for the full
model calculations. The results by switching off the contributions
from $\Lambda(1405)S_{01}$, $\Lambda(1670)S_{01}$,
$\Lambda(1520)D_{03}$, $\Lambda(1690)D_{03}$, $\Lambda$-pole, $u$-
and $t$-channel backgrounds are indicated explicitly by the legend.
}\label{poso}
\end{figure}

Combining the determined $C_R$ and $g_R$ factors, we derive the
ratios of the strengths ($\propto g_RC_R$) for the $S$- and $D$-wave
resonances in the reaction, which are
\begin{eqnarray}\label{i}
\mathcal{G}_{S_{01}(1405)}:\mathcal{G}_{S_{01}(1670)}\simeq-9:1,\\
\mathcal{G}_{D_{03}(1520)}:\mathcal{G}_{D_{03}(1690)}\simeq-7:1.
\end{eqnarray}
From the ratios, it is found that the $\Lambda(1405)S_{01}$ and
$\Lambda(1520)D_{03}$ govern the contributions of $S$ and $D$ waves,
respectively, in the reaction. The reversed signs in the two $S$
waves ($D$ waves) indicates that they have destructive interferences
each other.

It should be mentioned that our analysis suggests a much weaker
contribution of $\Lambda(1670)S_{01}$ in the reaction than that
derived from the symmetry quark model. The coupling strength
parameter, $C_R$, is about $1/3$ of that derived in the
SU(6)$\otimes$O(3) limit. The weaker contribution of
$\Lambda(1670)S_{01}$ might be explained by the configuration mixing
between $\Lambda(1405)S_{01}$ and
$\Lambda(1670)S_{01}$~\cite{Zhong:2009}. An \emph{et al.} also
predicted the existence of configuration mixings within the
$\Lambda(1405)S_{01}$ and $\Lambda(1670)S_{01}$ by analyzing the
decay properties of $\Lambda(1405)S_{01}$~\cite{An:2010wb}. On the
other hand, to well describe the data, it is needed a large
amplitude of $\Lambda(1520)D_{03}$ in the reaction, which is about a
factor of 2.5 larger than that derived in the SU(6)$\otimes$O(3)
limit (i.e., $C_{D_{03}(1520)}\simeq2.5 $ ). This means that we
underestimate the couplings of $\Lambda(1520)D_{03}$ to $\bar{K}N$
and/or $\pi\Sigma$ in the SU(6)$\otimes$O(3) limit for some reasons.

The dominant roles of $\Lambda(1405)S_{01}$ and
$\Lambda(1520)D_{03}$ in the reaction can be obviously seen in the
cross section, differential cross sections and $\Sigma^0$
polarizations. Switching off the contribution of
$\Lambda(1405)S_{01}$ or $\Lambda(1520)D_{03}$, the cross section,
differential cross sections and polarizations change dramatically
(see Figs. \ref{csa}-\ref{poso}). The cross sections in the low
energy region $P_{K^-}\lesssim 300$ MeV/c ($W\lesssim 1.49$ GeV) are
sensitive to the mass of $\Lambda(1405)S_{01}$. From Figs. \ref{csa}
and \ref{spi} it is seen that $\Lambda(1520)D_{03}$ plays a crucial
role in the low energy region. Around $P_{K^-}$ = 400 MeV/c
($W\simeq 1.52$ GeV), $\Lambda(1520)D_{03}$ is responsible for the
sharp resonant peak in the total cross section. The tail of the
partial cross section of $\Lambda(1520)D_{03}$ can extend to the low
energy region $P_{K^-}\sim 300$ MeV/c ($W\sim 1.49$ GeV), and to the
higher energy region $P_{K^-}\sim 750$ MeV/c ($W\sim1.68$ GeV),
which is consistent with the analysis of \cite{Bouzasa;2008}.

Although the contributions of $\Lambda(1670)S_{01}$ and
$\Lambda(1690)D_{03}$ are not as strong as those of
$\Lambda(1405)S_{01}$ and $\Lambda(1520)D_{03}$, their roles can be
seen around its threshold as well. If we switch off one of their
contributions, the differential cross sections and $\Sigma^0$
polarizations change significantly at $P_{K^-}\simeq 700\sim 800$
MeV/c ($W\simeq 1.65\sim 1.7$ GeV) (see Figs. \ref{spi}
and~\ref{poso}). From Fig.~\ref{csa}, it is found that the bump
structure around $P_{K^-}= 780$ MeV/c ($W\simeq 1.69$ GeV) in the
cross sections comes from the interferences between
$\Lambda(1670)S_{01}$ and $\Lambda(1690)D_{03}$. Turning off the
contributions of $\Lambda(1670)S_{01}$ and $\Lambda(1690)D_{03}$ at
the same time, the bump structure around $P_{K^-}= 780$ MeV/c
($W\simeq 1.69$ GeV) will disappear.

Both $u$- and $t$-channel backgrounds play crucial roles in the
reactions. Switching off $t$-channel contribution, the differential
cross sections are strongly overestimated at both forward and
backward angles, while the sign of the polarization is even changed.
The $u$-channel effects on the differential cross sections are also
can be obviously seen, if its contribution is switched off,
differential cross sections are overestimated significantly.

Finally, it should be pointed out that the $\Lambda$ pole also plays
an important role in the reaction. It has large effects on  both the
differential cross sections and $\Sigma^0$ polarizations in the
whole energy region what we considered, although it has negligible
effects on the total cross section. Switching off its contributions,
the differential cross sections and $\Sigma^0$ polarizations are
dramatically changed at both forward and backward angles (see
Figs.~\ref{spi} and ~\ref{poso}).

As a whole, the resonances $\Lambda(1405)S_{01}$ and
$\Lambda(1520)D_{03}$ play dominant roles in the reactions. Although
the contributions of $\Lambda(1670)S_{01}$ and $\Lambda(1690)D_{03}$
are much weaker than those of $\Lambda(1405)S_{01}$ and
$\Lambda(1520)D_{03}$, obvious evidence of $\Lambda(1670)S_{01}$ and
$\Lambda(1690)D_{03}$ in the reaction can be seen around their
threshold. The interferences between $\Lambda(1670)S_{01}$ and
$\Lambda(1690)D_{03}$ might be responsible for the bump structure
around $W= 1.69$ GeV in the cross section. There might exist
configuration mixings between $\Lambda(1405)S_{01}$ and
$\Lambda(1670)S_{01}$. The backgrounds also play crucial roles in
the reactions. The differential cross sections are sensitive to the
$\Lambda$ pole, although the total cross section is less sensitive
to it. The $u$- and $t$-channel backgrounds dramatically affects
both the angle distributions and total cross section.

\begin{widetext}
\begin{center}
\begin{figure}[ht]
\centering \epsfxsize=15.0 cm \epsfbox{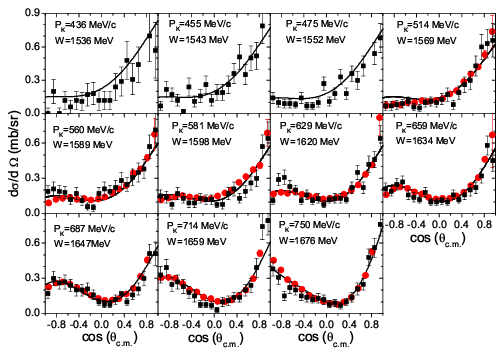} \epsfxsize=15.0 cm
\epsfbox{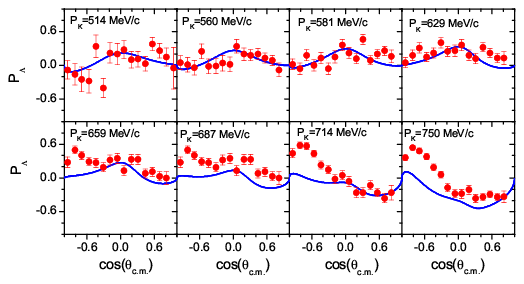} \caption{(Color online) Differential cross
sections (upper panel) and $\Lambda$ polarizations (lower panel) of
$K^-p\rightarrow \Lambda \pi^0$ compared with the data
from~\cite{Prakhov:2008} (solid circles), and \cite{armen:1970zh}
(squares). }\label{Lmbdapi}
\end{figure}
\end{center}
\end{widetext}

\begin{figure}[ht]
\centering \epsfxsize=8.0 cm \epsfbox{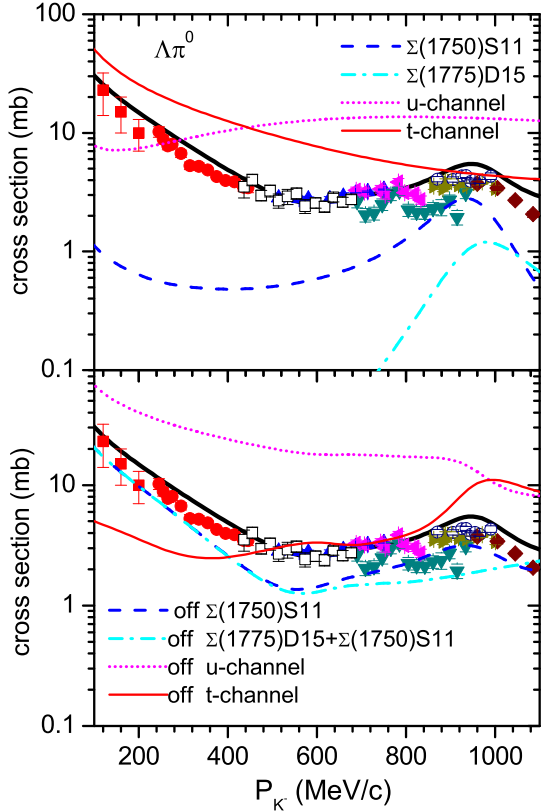} \caption{(Color
online) Cross section of the $K^-p\rightarrow \Lambda \pi^0$
process. The bold solid curves are for the full model calculations.
Data are from Refs. \cite{armen:1970zh} (open squares),
\cite{Mast:1976} ( solid circles), \cite{Ponte:1975bt}
(left-triangles), \cite{Jones:1974at} (right-triangles),
\cite{Baxter:1974zs} (down-triangles), \cite{Prakhov:2008}
(up-triangles), \cite{Cameron:1980nv} (open circles),
\cite{Kim:1965} (solid squares), \cite{Conforto:1975nw} (solid
diamonds). In the upper panel, exclusive cross sections for
$\Sigma(1750)$, $\Lambda(1775)$, $t$ channel, and $u$ channel are
indicated explicitly by the legends in the figures. In the lower
panel, the results by switching off the contributions of
$\Sigma(1750)$, $\Lambda(1775)$, $t$ channel, and $u$ channel are
indicated explicitly by the legends in the figures.}\label{cLmbdapi}
\end{figure}

\begin{figure}[ht]
\centering \epsfxsize=8.0 cm \epsfbox{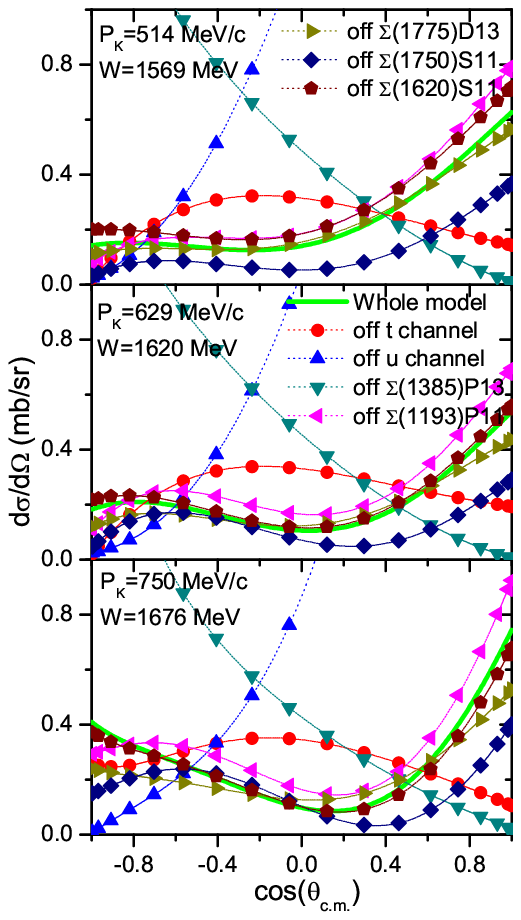} \caption{(Color
online) Effects of backgrounds and individual resonances on the
differential cross sections at three energies for the
$K^-p\rightarrow \Lambda \pi^0$ process. The bold solid curves are
for the full model calculations. The results by switching off the
contributions from $\Sigma(1193)$, $\Sigma(1385)$, $\Sigma(1620)$,
$\Sigma(1750)$, $\Sigma(1775)$, $u$- and $t$-channel backgrounds are
indicated explicitly by the legend.}\label{dfLmbdapi}
\end{figure}

\begin{figure}[ht]
\centering \epsfxsize=8.0 cm \epsfbox{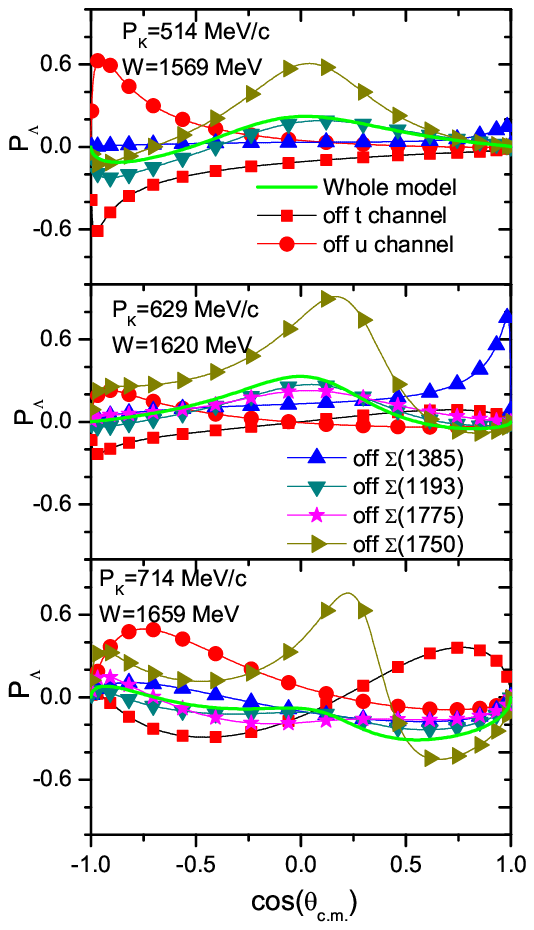} \caption{(Color
online) Effects of backgrounds and individual resonances on the
$\Lambda$ polarizations at three energies for the $K^-p\rightarrow
\Lambda \pi^0$ process. The bold solid curves are for the full model
calculations. The results by switching off the contributions from
$\Sigma(1193)$, $\Sigma(1385)$, $\Sigma(1620)$, $\Sigma(1750)$,
$\Sigma(1775)$, $u$- and $t$-channel backgrounds are indicated
explicitly by the legend. }\label{pLmbdapi}
\end{figure}

\subsection{$K^-p\rightarrow\Lambda \pi^0$}

In this reaction, the intermediate states of $s$ channel can only be
hyperons with isospin $I=1$ (i.e., $\Sigma$ hyperons) for the
isospin selection rule. Thus, this reaction provides us a rather
clear place to study the $\Sigma$ resonances. The low-lying $\Sigma$
resonances classified in the quark model are listed in Tab.
\ref{Resonance}. From the table, we see that in the $n=0$ shell
there are two $P$ waves: $\Sigma(1193)P_{11}$ and
$\Sigma(1385)P_{13}$. While in the $n=1$ shell, there exist three
$S_{11}$ waves: $[70,^210]S_{11}$, $[70,^28]S_{11}$ and
$[70,^48]S_{11}$; three $D_{13}$ waves: $[70,^210]D_{13}$,
$[70,^28]D_{13}$ and $[70,^48]D_{13}$; and one $D_{15}$ wave:
$[70,^48]D_{15}$. In these resonances only two states
$\Sigma(1670)D_{13}$ and $\Sigma(1775)D_{15}$ are well-established
\cite{PDG}. According to the classifications of the constituent
quark model, they correspond to the representations
$[70,^28,1,\textbf{1}]$ and $[70,^48,1,\textbf{1}]$, respectively.
Although, many $\Sigma$ resonances have been listed in PDG book
\cite{PDG}, their properties are still controversial. The $\Sigma$
spectroscopy is far from being established. In the present work, we
have carefully analyzed the new data of $K^-p\rightarrow\Lambda
\pi^0$. Our results compared with the data have been shown in
Figs.~\ref{Lmbdapi} and~\ref{cLmbdapi}.  From these figures, it is
seen that the low energy reaction $K^-p\rightarrow\Lambda \pi^0$ can
be well described with the parameters (See Tab.~\ref{param C})
determined by fitting 252 data points of the differential cross
sections and $\Lambda$ polarizations from Ref.~\cite{Prakhov:2008},
except some differences between theoretical results and the
observations of the $\Lambda$ polarizations in the higher energy
region $P_{K^-}\gtrsim 650$ MeV/c ($W\gtrsim 1.63$ GeV). The
$\chi^2$ per datum point is about $\chi^2/N=5.3$.

According to the determined $g_R$ and $C_R$ factors, we derive the
strength ratios between the $S$-wave resonances, which are
\begin{eqnarray}
\mathcal{G}_{S_{11}[70,^28]}:\mathcal{G}_{S_{11}[70,^48]}:\mathcal{G}_{S_{11}[70,^210]}\simeq-1:4:-1.
\end{eqnarray}
It indicates the dominant contributions of $S_{11}[70,^48]$ in the
$S$ waves. While, the derived strength ratios between the $D_{13}$
waves are
\begin{eqnarray}
\mathcal{G}_{D_{13}[70,^28]}:\mathcal{G}_{D_{13}[70,^48]}:\mathcal{G}_{D_{13}[70,^210]}\simeq5:-4:-5.
\end{eqnarray}
It is shown that these $D$-wave resonances with $J^P=3/2^-$ have
comparable contributions to the reaction.

From Tab.~\ref{Resonance}, it is found that the low-lying three
$S$-wave resonances $[70,^210]S_{11}$, $[70,^28]S_{11}$ and
$[70,^48]S_{11}$ are not established at all. According to the
predictions of the traditional quark model, the masses of the S-wave
$\Sigma$ resonances should be larger than $1.6$ GeV, however, the
PDG has listed some states with mass less than $1.6$ GeV, such as
$\Sigma(1480)$, $\Sigma(1560)$ and $\Sigma(1580)$. We carefully
analyze the differential cross sections, cross sections and
$\Lambda$ polarizations of the reaction $K^-p\rightarrow\Lambda
\pi^0$. We have found that the $[70,^48]S_{11}$ should have a mass
of $\sim 1770$ MeV, and a width of $\sim 90$ MeV. This state has a
significant contribution to the reaction in the whole resonance
regions what we considered. Both the mass and width of this
resonance are consistent with the 3-star resonance
$\Sigma(1750)1/2^-$ in PDG. Switching off its contributions, from
Figs. \ref{cLmbdapi}-\ref{pLmbdapi} we find that the cross sections
are obviously underestimated, and the shapes of the differential
cross sections and polarizations change dramatically. Furthermore,
we find that the resonance $[70,^28]S_{11}$ with a mass of $\sim
1631$ MeV and a width of $\sim 100$ MeV seems to be needed in the
reactions, with which the descriptions of the data around $W\simeq
1.6$ GeV is improved slightly (see Fig.~\ref{dfLmbdapi}). This
resonance is most likely to be the two-star state
$\Sigma(1620)1/2^-$ in PDG. The quark model classifications for
$\Sigma(1620)1/2^-$ and $\Sigma(1750)1/2^-$ suggested by us are
consistent with the suggestions in Ref.~\cite{Klempt:2009pi}. It
should be mentioned that no obvious evidence of $[70,^210]S_{11}$ is
found in the reaction.

In the $D$ waves, $\Sigma(1670)D_{13}$ and $\Sigma(1775)D_{15}$ are
two well-established states. According to the quark model
classifications, they correspond to the representations
$[70,^28,1,\textbf{1}]$ and $[70,^48,1,\textbf{1}]$, respectively.
Obvious roles of $\Sigma(1775)D_{15}$ can be found in the reaction.
From Fig.~\ref{cLmbdapi}, it is seen that the bump structure in the
cross section around $P_{K^-}= 950$ MeV/c ($W\simeq 1.77$ GeV) is
due to the interferences between $\Sigma(1775)D_{15}$ and
$\Sigma(1750)S_{11}$. The effects of $\Sigma(1775)D_{15}$ on the
differential cross sections can extend to the low energy region
$P_{K^-}\simeq 600$ MeV/c ($W\simeq 1.6$ GeV). Switching off the
contribution of $\Sigma(1775)D_{15}$, one can see that the
differential cross sections at very forward and backward angles
change significantly. No confirmed evidence for $\Sigma(1670)D_{13}$
and the other $D$-wave resonances is found in the reaction.

It should be mentioned that the ground $P$-wave state
$\Sigma(1385)P_{13}$ plays a crucial role in the reaction. Both the
differential cross sections and the $\Lambda$ polarizations are
sensitive to it. Switching off the contributions of
$\Sigma(1385)P_{13}$, one finds that the differential cross sections
and $\Lambda$ polarizations change dramatically (see the Figs.
\ref{dfLmbdapi} and \ref{pLmbdapi}). However, $\Sigma(1193)P_{11}$
has a small effect on the differential cross sections. Without
$\Sigma(1193)P_{11}$, the differential cross sections only enhance
slightly at forward angles.

$\Sigma(1660)P_{11}$ is a well established state in the energy
region what we considered, thus, we have analyzed its contributions
to the reaction $K^-p\rightarrow\Lambda \pi^0$. It should be pointed
out that we do not find any obvious evidence of $\Sigma(1660)P_{11}$
in the reaction. However, it should be mentioned that recently, Gao,
Shi and Zou had studied this reaction with an effective Lagrangian
approach as well, they claimed that their results clearly support
the existence of $\Sigma(1660)P_{11}$ in the
reaction~\cite{Gao:2012zh}.

Finally, it should be emphasized that the $u$- and $t$-channel
backgrounds play dominant roles in the reaction. Both of the two
channels not only are the main contributors to the total cross
sections, but also have large effects on the differential cross
sections and $\Lambda$ polarizations (see Figs. \ref{dfLmbdapi} and
\ref{pLmbdapi}).

In brief, the reaction $K^-p\rightarrow\Lambda \pi^0$ is dominated
by the $u$- and $t$-channel backgrounds and the ground $P$-wave
state $\Sigma(1385)P_{13}$. Furthermore, significant evidence of
$\Sigma(1775)D_{15}$ and $\Sigma(1750)S_{11}$ can be found in the
reaction. Some hints of $\Sigma(1620)1/2^-$ might exist in the
reaction, with which the descriptions of the data around $W\simeq
1.6$ GeV is improved slightly. It should be pointed out that no
confirmed evidence of the low mass resonances $\Sigma(1480)$,
$\Sigma(1560)$ and $\Sigma(1580)$ listed in PDG is found in the
$K^-p\rightarrow\Lambda \pi^0$ process.

\subsection{$K^-p\rightarrow \bar{K}^0 n$}

\begin{figure}[ht]
\centering \epsfxsize=8.6 cm \epsfbox{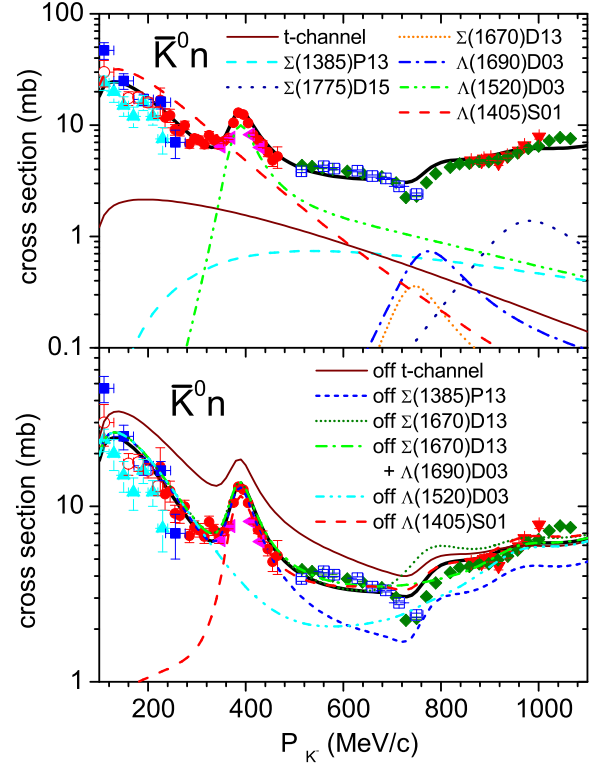} \caption{(Color
online) Cross section of the $K^-p\rightarrow \bar{K}^0n$ process.
The bold solid curves are the full model calculations. Data are from
Refs. \cite{Prakhov:2008} (open squares), \cite{Mast:1976} ( solid
circles), \cite{Berley:1996zh} (left-triangles), \cite{Jones:1974at}
(down-triangles), \cite{AlstonGarnjost:1977cs} (stars),
\cite{Kim:1965} (up-triangles), \cite{Evans:1983hz} (open circles),
\cite{Ciborowski:1982et} (solid squares). In the upper panel,
exclusive cross sections for $\Sigma(1385)$, $\Lambda(1405)$,
$\Lambda(1520)$, $\Lambda(1690)$, $\Sigma(1670)$, $\Sigma(1775)$ and
$t$ channel are indicated explicitly by the legends in the figures.
In the lower panel, the results by switching off the contributions
of $\Sigma(1385)$, $\Lambda(1405)$, $\Lambda(1520)$,
$\Lambda(1690)$, $\Sigma(1670)$, $\Sigma(1775)$ and $t$ channel are
indicated explicitly by the legends in the figures.}\label{ckn}
\end{figure}

Both the isospin-$1$ and isospin-$0$ intermediate hyperons can
contribute to the reaction $K^-p\rightarrow \bar{K}^0 n$. Thus, the
properties of $\Lambda$ and $\Sigma$ resonaces can be further
constrained and/or confirmed by the study of the $K^-p\rightarrow
\bar{K}^0 n$ process. We have analyzed the data of the low energy
reaction $K^-p\rightarrow \bar{K}^0 n$. Our results compared with
the data are shown in Figs.\ref{dfkn} and \ref{ckn}. From these
figures, it is found that the data are described fairly well within
our chiral quark model. However, we also notice that our theoretical
results might underestimate the differential cross sections at
forward angles in the higher energy region $P_{K^-}\gtrsim 560$
MeV/c ($W\gtrsim 1.59$ GeV).

\begin{widetext}
\begin{center}
\begin{figure}[ht]
\centering \epsfxsize=16.0 cm \epsfbox{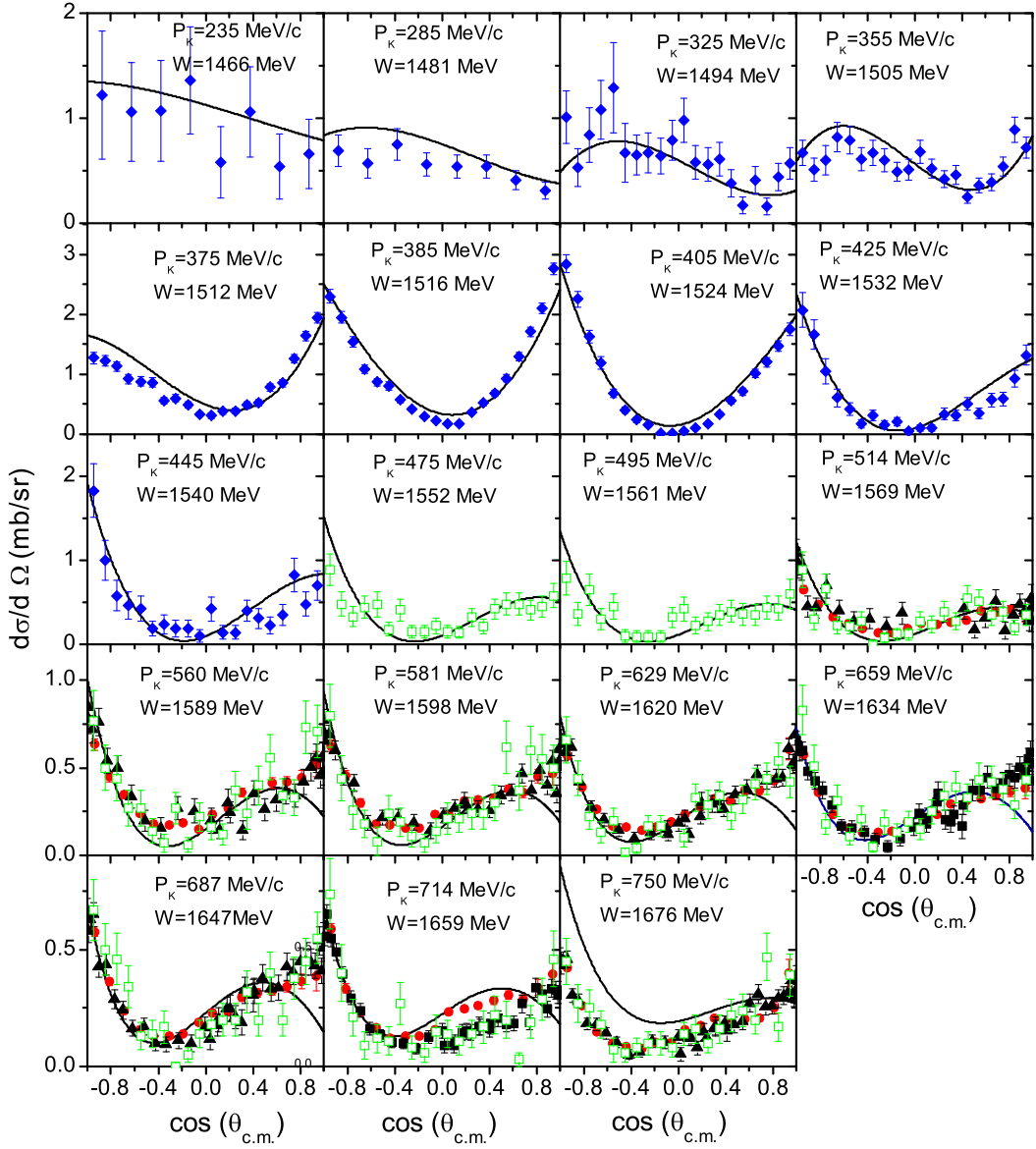}\caption{ (Color
online) Differential cross sections of $K^-p\rightarrow \bar{K}^0 n$
compared with the data are from~\cite{Mast:1976} (diamond),
\cite{Prakhov:2008} (solid circles), \cite{armen:1970zh} (open
squares), ~\cite{AlstonGarnjost:1977ct} (up triangles).}\label{dfkn}
\end{figure}
\end{center}
\end{widetext}

In the SU(6)$\otimes$O(3) symmetry limit, the couplings of
$\Lambda(1405)S_{01}$ and $\Lambda(1670)S_{01}$ to the $\bar{K}N$
should have the same value, however, the data favor a much weaker
coupling of $\Lambda(1670)S_{01}$ to $\bar{K}N$ than that of
$\Lambda(1405)S_{01}$, i.e., $|g_{\Lambda(1670)\bar{K}N}|\ll
|g_{\Lambda(1405)\bar{K}N}|$, which consists with our previous
analysis of the $K^-p\rightarrow \Sigma^0 \pi^0$ process, where a
very weak coupling of $\Lambda(1670)S_{01}$ to the $\bar{K}N$ is
also needed. The weak coupling of $\Lambda(1670)S_{01}$ to
$\bar{K}N$ was predicted with U$\chi $PT approach as
well~\cite{Oset:2001cn} . The configuration mixing between
$\Lambda(1670)S_{01}$ and $\Lambda(1405)S_{01}$ might explain the
weak coupling of $\Lambda(1670)S_{01}$ to
$\bar{K}N$~\cite{Zhong:2009}. Furthermore, we find that the data
favor a large coupling of $\Lambda(1520)D_{03}$ to $\bar{K}N$, which
indicates that we might underestimate the coupling of
$\Lambda(1520)D_{03}$ to $\bar{K}N$ in the SU(6)$\otimes$O(3) limit.
We also note that a larger coupling of $\Lambda(1520)D_{03}$ to
$\bar{K}N$ is needed in the $K^-p\rightarrow\Sigma^0\pi^0$ process.

According to the determined $C_R$ parameters and $g_R$ factors, the
ratios of the couplings of $S$-wave resonances to $\bar{K}N$ are
obtained:
\begin{eqnarray}
|g_{\Lambda(1405)\bar{K}N}|:|g_{\Lambda(1670)\bar{K}N}|:|g_{{S_{11}[70,^28]\bar{K}N}}|
:|g_{{S_{11}[70,^48]\bar{K}N}}|\nonumber\\
:|g_{{S_{11}[70,^210]\bar{K}N}}|\simeq 5.2:1.9:1:2.5:1.
\end{eqnarray}
From the ratios, it is seen that $\Lambda(1405)S_{01}$ dominates the
$S$-wave contributions in the reaction. While, for the $D$-wave
resonances, the ratios of their couplings to $\bar{K}N$ are
\begin{eqnarray}\label{f}
|g_{\Lambda(1520)\bar{K}N}|:|g_{\Lambda(1690)\bar{K}N}|:|g_{\Sigma(1670)\bar{K}N}|
:|g_{{D_{13}[70,^48]\bar{K}N}}|\nonumber\\:|g_{{D_{13}[70,^210]\bar{K}N}}|\simeq
9:3:2:1.
\end{eqnarray}
It shows that $\Lambda(1520)D_{03}$ dominates $D$-wave contributions
in the reaction.

Combing the ratios given in Eqs. (\ref{i})-(\ref{f}), we can easily
estimate some other important ratios:
\begin{eqnarray}
\left|\frac{g_{\Lambda(1405)\Sigma \pi}}{g_{\Lambda(1670)\Sigma
\pi}}\right|\simeq 3.8, \ & & \left|\frac{g_{\Lambda(1520)\Sigma
\pi}}{g_{\Lambda(1690)\Sigma \pi}}\right|\simeq 4.0.
\end{eqnarray}

\begin{figure}[ht]
\centering \epsfxsize=8.6 cm \epsfbox{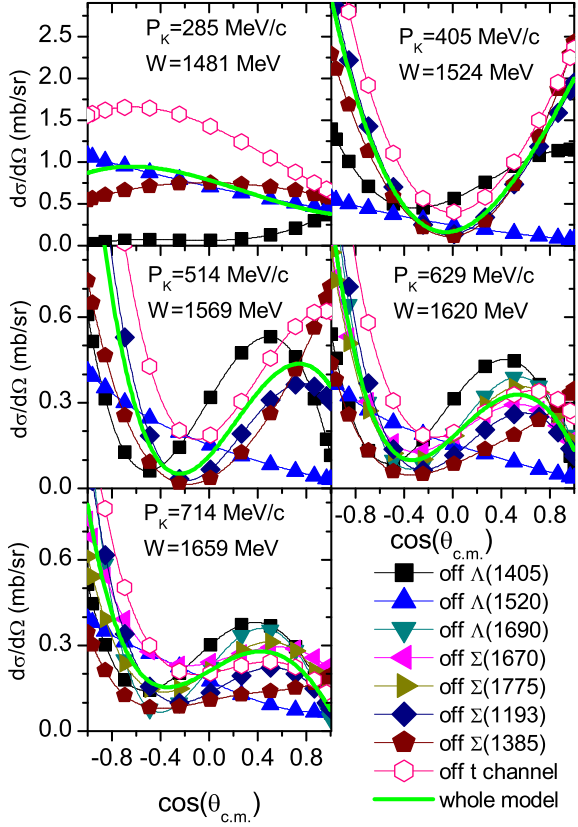} \caption{(Color
online) Effects of backgrounds and individual resonances on the
differential cross sections at five energies for the
$K^-p\rightarrow \bar{K}^0 n$ process. The bold solid curves are for
the full model calculations. The results by switching off the
contributions from $\Sigma(1193)$, $\Sigma(1385)$, $\Sigma(1670)$,
$\Sigma(1775)$, $\Lambda(1405)$, $\Lambda(1520)$ and $t$-channel
backgrounds are indicated explicitly by the legend. }\label{dfkno}
\end{figure}

In Figs.\ref{ckn} and \ref{dfkno}, we have shown the contributions
of the main partial waves to the cross sections and differential
cross sections in the reaction. From the figure, it is seen that
$\Lambda(1405)S_{01}$ dominates the reaction at low energies.
Switching off its contributions, we find that the cross sections at
$P_{K^-}< 400$ MeV/c ($W< 1.52$ GeV) are dramatically underestimated
(see Fig.~\ref{ckn}). The differential cross sections are sensitive
to $\Lambda(1405)S_{01}$ in the whole energy region what we
considered, although it has less effects on the total cross sections
at $P_{K^-}\gtrsim 400$ MeV/c ($W\gtrsim 1.52$ GeV).

The $D$-wave resonance $\Lambda(1520)D_{03}$ is crucial to the
reaction as well. It is responsible for the sharp peak at
$P_{K^-}\simeq 400$ MeV/c ($W\simeq 1.52$ GeV) in the cross section.
The strong effects of $\Lambda(1520)D_{03}$ on the reaction can
extend to the higher energy region $P_{K^-}\simeq 1000$ MeV/c
($W\simeq 1.79$ GeV). Switching off its contributions, the
differential cross sections change dramatically in a wide energy
region $P_{K^-}>300$ MeV/c ($W>1.49$ GeV). The bowl shape of the
differential cross section at $P_{K^-}\simeq 400$ MeV/c ($W\simeq
1.52$ GeV) is caused by the interferences between
$\Lambda(1520)D_{03}$ and $\Lambda(1405)S_{01}$.

Furthermore, from Figs.~\ref{ckn} and \ref{dfkno}, we can see slight
effects of $\Lambda(1690)D_{03}$, $\Sigma(1670)D_{13}$ and
$\Sigma(1775)D_{15}$ on the differential cross sections and cross
sections in the energy region $P_{K^-}\gtrsim 600$ MeV/c ($W\gtrsim
1.6$ GeV). Their contributions to the cross section are much smaller
than those of $\Lambda(1520)D_{03}$. The interferences between
$\Lambda(1690)D_{03}$ and $\Sigma(1670)D_{13}$ might be responsible
for the dip structure at $P_{K^-}\simeq 750$ MeV/c ($W\simeq 1.68$
GeV) in the cross section. If we switch off their contribution in
the reaction, this dip structure will disappear (see
Fig.~\ref{ckn}).

In the $s$-channel ground states, it is seen that
$\Sigma(1385)P_{13}$ plays an important role in the reaction.
Without it, the cross sections in the region $P_{K}\gtrsim 500$
MeV/c ($W\gtrsim 1.56$ GeV) are obviously underestimated, and the
differential cross sections are changed significantly. However, only
a small contribution of $\Sigma(1193)P_{11}$ to the reaction is seen
in the differential cross sections at $P_{K}\gtrsim 600$ MeV/c
($W\gtrsim 1.6$ GeV). Switching off it, the cross sections at
forward angles should be slightly underestimated.

In the reaction $K^-p\rightarrow \bar{K}^0 n$, the $t$-channel
background also plays a crucial role. Switching off it, we have
noted that the cross sections are overestimated significantly, and
the shapes of the differential cross sections change dramatically.
The $t$ channel has significant destructive interferences with
$\Lambda(1405)S_{01}$.

It should be mentioned that in the $K^-p\rightarrow \bar{K}^0 n$
process, we do not find any confirmed evidence of
$\Lambda(1670)S_{01}$, $\Sigma(1620)S_{11}$, $\Sigma(1750)S_{11}$,
$\Lambda(1600)P_{01}$ and $\Sigma(1660)P_{11}$. Furthermore, we do
not find any evidence of the low mass $\Sigma$ resonances
$\Sigma(1480)$, $\Sigma(1560)$ and $\Sigma(1580)$ listed in PDG.

Summarily, $\Lambda(1405)$ and $\Lambda(1520)D_{03}$ and $t$-channel
background govern the $K^-p\rightarrow \bar{K}^0 n$ process at the
low energy regions. The ground $P$-wave state $\Sigma(1385)P_{13}$
also plays an important role in the reaction. Furthermore, some
evidence of $\Lambda(1690)D_{03}$, $\Sigma(1670)D_{13}$ and
$\Sigma(1775)D_{15}$ are seen in the reaction. The interferences
between $\Lambda(1690)D_{03}$ and $\Sigma(1670)D_{13}$ might be
responsible for the dip structure at $P_{K^-}\simeq 750$ MeV/c
($W\simeq 1.68$ GeV) in the cross section.

\section{summary}\label{SUM}

In this work, we have carried out a combined study of the reactions
$K^-p\rightarrow\Lambda \pi^0$, $\Sigma^0 \pi^0$ and $\bar{K}^0 n$
in a chiral quark model. Good descriptions of the observations have
been obtained at low energies. In these processes, the roles of the
low-lying strangeness $S$=$-1$ hyperon resonances are carefully
analyzed, and the properties of some hyperon resonances are derived.

By studying the $K^-p\rightarrow\Lambda \pi^0$ process, we find some
significant evidence of the $\Sigma$ resonance $[70,^48]S_{11}$.
Both its mass and width are consistent with the 3-star resonance
$\Sigma(1750)1/2^-$ in PDG. Furthermore, we find that some hints of
$[70,^28]S_{11}$ might exist in the reaction, its mass and width are
consistent with the 2-star resonance $\Sigma(1620)1/2^-$ in PDG.
Obvious evidence of the $D$-wave resonance $\Sigma(1775)D_{15}$ is
also found in the reaction. The bump structure in the cross sections
around $W= 1.77$ GeV is due to the interferences between
$\Sigma(1775)D_{15}$ and $\Sigma(1750)S_{11}$.

In the $K^-p\rightarrow\Sigma^0 \pi^0$, $\bar{K}^0 n$ processes,
both $\Lambda(1405)S_{01}$ and $\Lambda(1520)D_{03}$ play crucial
roles. The observations of the two processes are sensitive to
$\Lambda(1405)S_{01}$ and $\Lambda(1520)D_{03}$.
$\Lambda(1520)D_{03}$ is responsible for the sharp peak in the cross
sections around $P_{K^-}=400$ MeV/c ($W\simeq 1.52$ GeV).
$\Lambda(1520)D_{03}$ has a larger coupling to $\bar{K}N$ than that
derived in the SU(6)$\otimes$O(3) limit. Furthermore, in both of the
processes $K^-p\rightarrow\Sigma^0 \pi^0$ and $\bar{K}^0 n$, a weak
coupling of $\Lambda(1670)S_{01}$ to $\bar{K}N$ is needed, which
might be explained by the configuration mixing between
$\Lambda(1670)S_{01}$ and $\Lambda(1405)S_{01}$.

Some evidence of $\Lambda(1670)S_{01}$ and $\Lambda(1690)D_{03}$
around their threshold is found in the $K^-p\rightarrow\Sigma^0
\pi^0$ process. The interferences between $\Lambda(1670)S_{01}$ and
$\Lambda(1690)D_{03}$ might be responsible for the bump structure
around $W= 1.68$ GeV in the cross section.

Slight contributions from $\Lambda(1690)D_{03}$,
$\Sigma(1670)D_{13}$ and $\Sigma(1775)D_{15}$ are found in the
$K^-p\rightarrow\bar{K}^0 n$ process. The interferences between
$\Lambda(1690)D_{03}$ and $\Sigma(1670)D_{13}$ might be responsible
for the dip structure at $W\simeq 1.68$ GeV in the cross section.

The $u$- and $t$-channel backgrounds play crucial roles in the
reactions $K^-p\rightarrow\Sigma^0 \pi^0$, $\Lambda \pi^0$. The
important role of $u$ channel in these reactions is also predicted
with the B$\chi$PT \cite{Bouzasa;2008} and U$\chi$PT approaches
\cite{Oller:2006jw,Oller:2005ig}. In $K^-p\rightarrow\bar{K}^0 n$,
there are no $u$-channel contributions, while the $t$ channel is
crucial to give the correct shapes of the differential cross
sections. The important role of $t$ channel is also found in the
$KN$ and $\pi N$ processes \cite{Wu:2007fc}.

The $s$-channel Born term plays an important role in the reactions.
For the $K^-p\rightarrow\Lambda \pi^0$ process, both the
differential cross sections and the $\Lambda$ polarizations are
sensitive to $\Sigma(1385)P_{13}$. For the $K^-p\rightarrow\Sigma^0
\pi^0$ process, the $\Lambda$ pole has large effects on the
differential cross sections and $\Sigma^0$ polarizations in the
whole energy region what we considered, although it has little
effects on the total cross section. And for the
$K^-p\rightarrow\bar{K}^0 n$ process, $\Sigma(1385)P_{13}$ has
obvious effects on the cross sections and the differential cross
sections.

It should be mentioned that no evidence of the low mass resonances
$\Sigma(1480)$, $\Sigma(1560)$ and $\Sigma(1580)$ listed in PDG are
found in the $K^-p\rightarrow\Lambda \pi^0$, $\Sigma^0 \pi^0$ and
$\bar{K}^0 n$ processes.

In this work we have only analyzed the $K^-p$ scattering to neutral
final states. In our future work, we will carry out a systemical
study of the reactions $K^-p$ scattering to charged final states,
$K^-p\rightarrow \Sigma^{\pm}\pi^{\mp},K^-p$.  Of course, we expect
high precision measurements of these reactions are to be performed
at J-PARC in future experiments.


\section*{  Acknowledgements }
This work is supported, in part, by the National Natural Science
Foundation of China (Grants No. 11075051 and No. 11035006), Chinese
Academy of Sciences (KJCX2-EW-N01), Program for Changjiang Scholars
and Innovative Research Team in University (PCSIRT, Grant No.
IRT0964), the Program Excellent Talent Hunan Normal University, and
the Hunan Provincial Natural Science Foundation (Grants No. 11JJ7001
and No. 13JJ1018).

\end{document}